\documentclass[11pt]{article}
\usepackage{graphics,epsfig,amsfonts,amsmath,amssymb,bbm}
\usepackage{float}
\usepackage{relsize}
\usepackage{accents}
\usepackage{mathrsfs}
\usepackage{graphicx}
\usepackage{soul, color}
\usepackage[normalem]{ulem}
\usepackage{enumitem}
\usepackage[T1]{fontenc}
\usepackage{array}
\usepackage{makecell}
\usepackage{hyperref}
\usepackage[framed]{matlab-prettifier}

\newcolumntype{x}[1]{>{\centering\let\newline\\\arraybackslash\hspace{0pt}}p{#1}}
\begin{document}
 \begin{flushleft}

{\Large
\textbf{{Uncertainty and Sensitivity Analyses Methods for Agent-Based Mathematical Models: An Introductory Review }}}

\vskip1cm
Sara Hamis$^{1,2}$, 
Stanislav Stratiev$^{3}$, 
Gibin G Powathil$^{2}$.

\vskip.4cm
$\bf{^1}$ School of Mathematics and Statistics, University of St Andrews, St Andrews KY16 9SS, Scotland. \\
$\bf{^2}$ Department of Mathematics, College of Science, Swansea University, Swansea, SA1 8EN, United Kingdom.\\
$\bf{^3}$ Department of Physics, College of Science, Swansea University, Swansea, SA2 8PP, United Kingdom.\\

 \end{flushleft}
 \vskip.4cm

\begin{abstract}
\vspace{1cm}
\noindent 
Multiscale, agent-based mathematical models of biological systems are often associated with model uncertainty and sensitivity to parameter perturbations.  
Here, three uncertainty and sensitivity analyses methods, that are suitable to use when working with agent-based models, are discussed. These methods are namely Consistency Analysis, Robustness Analysis and Latin Hypercube Analysis. This introductory review discusses origins, conventions, implementation and result interpretation of the aforementioned methods. 
Information on how to implement the discussed methods in \textsf{MATLAB} is included.

\end{abstract}
 

\section{Introduction}
Mathematical models of biological systems are abstractions of highly complex reality. It follows that parameters used in such models often are associated with some degree of uncertainty, where the uncertainty can be derived from various origins. Epistemic uncertainty refers to uncertainty resulting from limited knowledge about the biological system at hand, whilst aleatory uncertainty stems from naturally occurring stochasticity, intrinsic to biological systems \cite{Read2012, Spartan2013, Lin2016}. 
Model parameters may thus be naturally stochastic, theoretically unknown, and unfeasible or impossible to measure precisely (or at all). 
Further magnifying the contributions of uncertainty in mathematical models of biological systems, in particular, is the fact that \textit{one} parameter in the mathematical model may correspond to \textit{a multitude} of underlying biological mechanisms and factors in the real, biological system. This is especially true for minimal parameter models, {\em i.e.} mathematical models that aspire to be as non-complex as possible whilst still capturing all biological details of interest \cite{Gevertz2018}.  
In this review, we focus our attention on uncertainty and sensitivity analyses techniques that are suitable for use with \textit{agent-based} models. A mathematical, agent-based model comprises several distinct \textit{agents} that may interact with each other and their environment. In an agent-based tumour model, for example, an agent typically corresponds to one tumour cell or a group of tumour cells. This naturally allows for heterogeneity amongst tumour cells, which is useful as tumour heterogeneity is associated with many complications involved in modelling (and treating) solid tumours. Accordingly, many modellers in the field of mathematical oncology choose to work with agent-based models \cite{Rejniak2011}.

There already exist multiple method papers that describe \textit{how} to perform uncertainty and sensitivity analyses when working with agent-based models, authors Alden {\em et al.} even provide a free R-based software package (Spartan \cite{Spartan2013}) that enables the user to perform different such methods, including the three methods discussed in this review. However, as these methods have been developed across multiple research fields, both inside and outside of the natural sciences, it is difficult to find one comprehensive review that discusses not only \text{how} to perform these methods, but also {\it where} these methods come from, and {\it why} certain conventions are proposed and/or used. 
To this end, we have in this review gathered such information for three uncertainty and sensitivity analyses techniques, namely Consistency Analysis, Robustness Analysis and Latin Hypercube Analysis. 
Our aim is that this will allow the reader to better evaluate uncertainty and sensitivity analyses presented by other authors, and encourage the reader to consider performing these methods when suitable. 
\\

In order to understand the impact that parameter uncertainty and parameter perturbations have on results produced by a mathematical model, uncertainty and sensitivity analyses can be used. 
A mathematical model that comprises a set of uncertain model parameters (or \textit{inputs}), is able to produce a range of possible responses (or \textit{outputs}). 
%
\textit{Uncertainty analysis} assesses the range of these outputs overall, and provides information regarding how certain (or uncertain) we should be with our model results, and the conclusions that we draw from them \cite{Blower1994}. 
\textit{Sensitivity analysis} describes the relationship between uncertainty in inputs and uncertainty in outputs. It can be used to identify which sources of input uncertainty ({\em i.e.} which model parameters) significantly influence the uncertainty in the output and, equally importantly, which do not \cite{Blower1994}. 
Assessing how sensitive the output is to small input perturbations is a healthy way to scrutinise our mathematical models \cite{Ligmann-Zielinska2014}. Moreover, for a well-formulated model, knowledge regarding how input uncertainty influences output uncertainty can yield insight into the biological system that has not yet been empirically observed  \cite{Spartan2013}. Furthermore, if the uncertainty in some input parameter is shown to not affect output uncertainty, the modeller may consider fixing that parameter, and thus reducing model complexity in accordance with a minimal-parameter modelling approach. 
In {\it local} sensitivity analysis techniques, model parameters (inputs) are perturbed one at a time whilst other parameters remain fixed at their calibrated value. In {\it global} sensitivity analysis techniques, all model parameters are simultaneously perturbed \cite{Charzynska2012}. 
\\

There exist several sensitivity and uncertainty analyses techniques, but here we will focus on three such techniques that are suitable to use in conjunction with agent-based mathematical models.  
These techniques are namely Consistency Analysis, Robustness Analysis and Latin Hypercube Analysis, which all answer important, and complementary, questions about mathematical models and their corresponding {\it in silico} responses \cite{Read2012, Spartan2013}.%
\\

\includegraphics[scale=0.375]{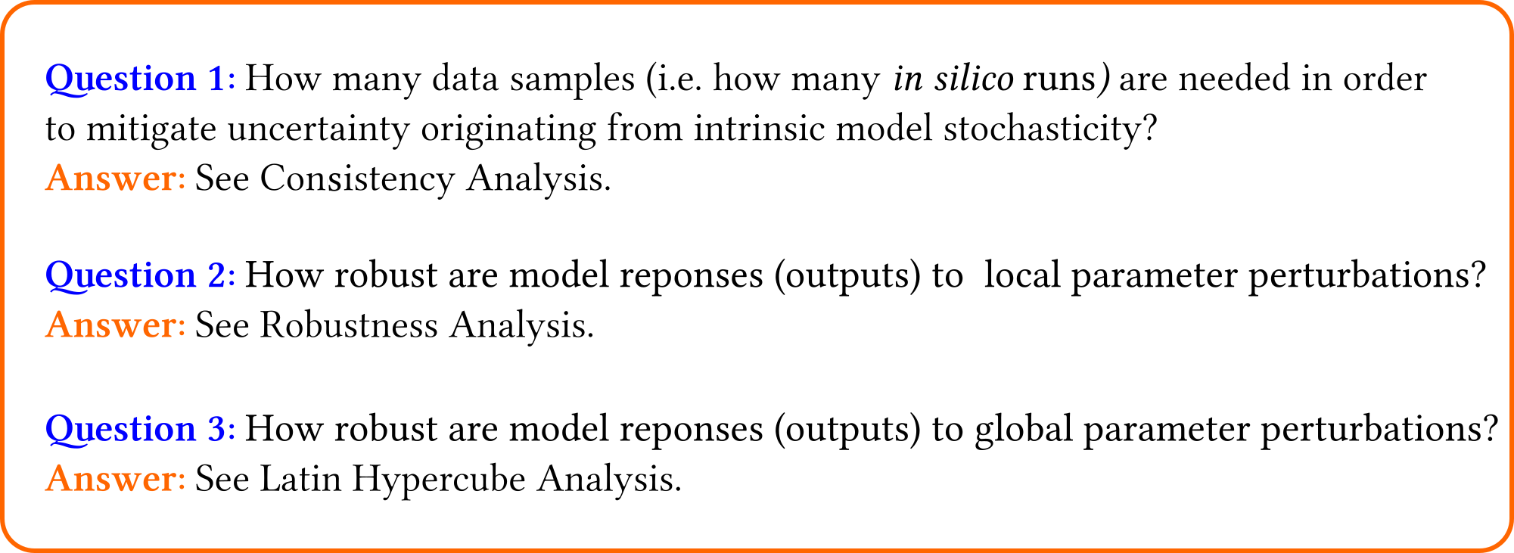}

\vspace{0.5cm}

Note that Consistency Analysis is only meaningful when analysing models with stochastic variables. 
\\

The statistical techniques described in this review have been developed and applied across multiple academic disciplines, both inside and outside of the natural sciences. Consequently, terminology and notations vary in the literature. The aim of this review is to combine pertinent literature from various academic fields whilst keeping terminology and mathematical notations consistent, unambiguous and tailored towards a mathematical and scientific audience. Therefore, when needed, certain algorithms from the literature are here reformulated into expressions that a mathematician would consider to be conventional. 
This review is intended to provide gentle, yet comprehensive, instructions to the modeller wanting to perform uncertainty and sensitivity analyses on agent-based models. 
Thorough directions on how to perform Consistency Analysis (Section \ref{sec:sa_ca}), Robustness Analysis (Section \ref{sec:sa_ra}) and Latin Hypercube Sampling and Analysis (Section \ref{sec:sa_lhc}) are provided.  Consistency Analysis utilises \textit{the A-measure of stochastic superiority}, which is therefore discussed in Section \ref{sec:sa_atest}. 
Throughout this review, we have included some historical information that elucidates why certain statistical conventions are used. Each section also contains pictorial, step-by-step instructions on how to perform the aforementioned techniques.  
Worked examples of all methods discussed in this review are provided in Section \ref{sec:workedexample}. These worked examples use \textit{in silico} data produced in a previous agent-based, multiscale, mathematical oncology study \cite{DDR}.

\subsubsection*{Methods outside the scope of this review}
Note that there exist other uncertainty and sensitivity analyses techniques, suitable for agent-based models, that are outside the scope of this review. 
For example, Bayesian inference is a statistical technique that uses Bayes theorem, prior beliefs and observed data to infer input parameter values and their associated uncertainties \cite{Lambert2018}. When this inference is difficult to calculate, computational methods such as Approximate Bayesian Computation (ABC) \cite{Liepe2014} or, if the model is highly computationally expensive, Approximate Approximate Bayesian Computation (AABC) \cite{Buzbas2015} can be used to perform inference on input model parameters. 
The Sobol method is a variance-based sensitivity analysis method that, using Monte-Carlo multidimensional integration, allows for the evaluation of each individual parameter's fractional contribution to the output variance \cite{Zhang_Sobol2015}. Another global, variance based sensitivity analysis method that enables quantification of the input parameter's fractional contributions to the output variance is Fourier amplitude sensitivity testing (FAST) which uses an underlying algorithm that involves Fourier decomposition \cite{Saltelli1998}. Although the Sobol method and FAST use different underlying techniques, they both allow us to say that ``input parameter $r_i$ contributes $R_i$ percent to the output variance''.  
Newer sensitivity analysis methods and tools tend to take advantage of the current abundance of computing power. The recently introduced MASSIVE (Massively parallel Agent-based Simulations and Subsequent Interactive Visualization-based Exploration) methodology, for example, combines parallel computing and interactive data visualisation to produce a graphical user interface that provides an overview of input-output relations for a broad input parameter range \cite{Niida2019}.

\section{The A-measure of stochastic superiority}
\label{sec:sa_atest}

\subsection{The Common Language Statistics}

In 1992, McGraw and Wong introduced the \textit{common language statistics} ($CL$) as an intuitive way to compare two distributions of data \cite{McGrawWong}.  
The $CL$ was initially introduced as a tool to compare data from normal distributions, but was later on approximated for use on any continuous distributions. 
The $CL$ describes the probability that a random data sample from one of the distributions is greater than a random data sample from the other distribution. 
For example, if we have two continuous data distributions $B$ and $C$, and we are comparing the distributions with respect to some variable $X$, then the $CL$ is simply given by
\begin{equation}
CL_{BC}(X)=P(X_B>X_C),
    \label{eq:cl}
\end{equation}

where standard probability notations have been used so that $P(X_B>X_C)$ denotes the probability that a random data sample $X_B$ from distribution $B$ is greater than a random data sample $X_C$ from distribution $C$ \cite{McGrawWong}. 
Thus the subscript of $X$ here signifies the distribution from which the data sample $X$ was taken.

\subsection{The A-measure of stochastic superiority}
The $CL$ was developed to compare continuous data distributions, but Vargha and Delaney \cite{VarghaDelaney} introduced the \textit{A-measure of stochastic superiority} (or A-measure for short) as a generalisation of the $CL$ that can directly be applied to compare both continuous and discrete distributions of variables that are at least ordinally scaled. 
When comparing two distributions $B$ and $C$, with respect to the variable $X$, the A-measure $A_{BC}(X)$ is given by 

\begin{equation}
A_{BC}(X)=P(X_B > X_C) + 0.5P(X_B=X_C),
    \label{eq:ameasure}
\end{equation}

where $P(X_B=X_C)$ denotes the probability that a random data sample from distribution $B$ is equal to a random data sample from distribution $C$. 
By comparing Equations \ref{eq:cl} and \ref{eq:ameasure}, it is clear that in the continuous case, where $P(X_B=X_C)=0$, the A-measure reduces to the $CL$. 
\\

If two distributions that are identical with respect to the variable $X$ are compared, then $P(X_B >X_C) = P(X_C > X_B)$ and we say that the distributions $B$ and $C$ are \textit{stochastically equal} with respect to the variable $X$.
On the other hand, if $P(X_B >X_C) > P(X_C > X_B)$, then we say that the distribution $B$ is \textit{stochastically greater than} distribution $C$, and accordingly, that distribution $C$ is \textit{stochastically smaller than} distribution $B$ \cite{VarghaDelaney}. 
If distribution $B$ is stochastically greater than distribution $C$ with respect to the variable $X$, it simply occurs more often that the sample $X_B$ is greater than the sample $X_C$ when two random samples $X_B$ and $X_C$ are compared. 
These definitions of stochastic relationships (\textit{stochastically equal to, stochastically greater than, stochastically smaller than}), used by Vargha and Delaney \cite{VarghaDelaney}, amongst others, are weaker than definitions used by some other authors, but sufficient and appropriate for our current purposes: comparing distributions of discrete data samples produced by {\it in silico} simulations based on stochastic, agent-based mathematical models. 
\\

When comparing two samples $X_B$ and $X_C$, the possible outcomes are \textit{(i)} that $X_B$ is greater than $X_C$, \textit{(ii)} that $X_B$ is equal to $X_C$ and \textit{(iii)} that $X_B$ is smaller than $X_C$. These three possible outcomes must sum up to one so that,

\begin{equation}
    P(X_B>X_C)+P(X_B=X_C)+P(X_C>X_B)=1.
    \label{eq:probisone}
\end{equation}

In the continuous case, $P(X_B=X_C)=0$ as previously stated, and thus it follows that 

\begin{equation}
    P(X_C>X_B)=1-P(X_B>X_C), \quad \text{for continuous distributions,}
\end{equation}

and thus it suffices to know only one of the values $P(X_B>X_C)$ or $P(X_C>X_B)$ in order to determine the stochastic relationship between the distributions $B$ and $C$ with respect to $X$. 
\\

\begin{itemize}
\item[$\blacktriangleright$] \textit {For example: if $P(X_B>X_C)=0.4$, then it is clear that $P(X_C>X_B)=0.6$ and thus that $P(X_B>X_C)< P(X_C>X_B)$, or equivalently, that distribution $B$ is stochastically smaller than distribution $C$.}
\end{itemize}

However, in the discrete case, $P(X_B=X_C)$ is not generally equal to zero and therefore,

\begin{equation}
    \label{eq:p_disc}
    P(X_C>X_B)=1-P(X_B>X_C)-P(X_B=X_C)  \quad \text{for discrete distributions.}
\end{equation}

Consequently, one single value $P(X_B>X_C)$ or $P(X_C>X_B)$ alone can generally not be used to determine the stochastic relationship between the distributions $B$ and $C$. 
\\

\begin{itemize}
\item[$\blacktriangleright$] \textit{For example: if, again, $P(X_B>X_C)=0.4$, it follows that $P(X_C>X_B)=0.6-P(X_B=X_C)$. This does not give us enough information to determine the stochastic relationship between the two distributions $B$ and $C$.}
\end{itemize}

In order to proceed to compare the distributions $B$ and $C$ in this case, the \textit{stochastic difference} $\delta$ is introduced, where $\delta$ is given by 

\begin{equation}
\delta = P(X_B>X_C) - P(X_C>X_B), \quad \delta \in [-1,1].
\label{eq:stochdiff}
\end{equation}

Via a linear transformation, the \textit{transformed stochastic difference}, $\delta' \in [0,1]$, can be obtained using Equation \ref{eq:p_disc} so that

\begin{multline*}
\delta' = \frac{\delta + 1}{2} = \frac{P(X_B>X_C)-P(X_C>X_B) + 1}{2} = \\
= \frac{P(X_B>X_C) - \big( 1-P(X_B>X_C)-P(X_B=X_C) \big) + 1}{2} =  \\
= P(X_B>X_C)+ 0.5P(X_B=X_C) =A_{BC}(X),
\end{multline*}
\begin{equation}
\end{equation}

from which we can see that the A-measure, $A_{BC}(X)$ (Equation \ref{eq:ameasure}), measures the stochastic difference between $P(X_B>X_C)$ and $P(X_C>X_B)$ under a linear transformation \cite{VarghaDelaney}. 
\\

\begin{figure}[H]
    \centering
    \includegraphics[scale=0.28]{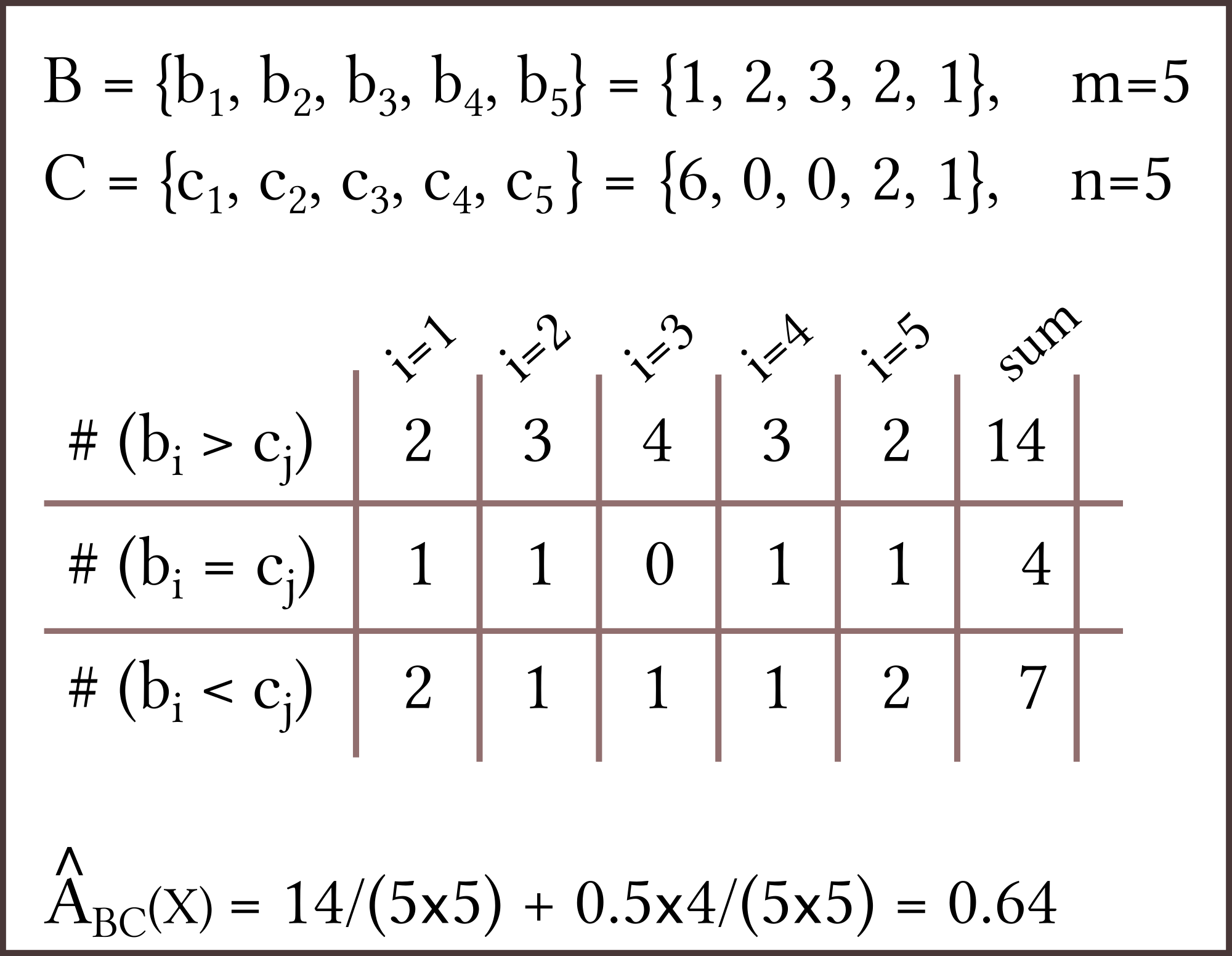}
    \caption{Using Equation \ref{eq:sa_peAm_count} to compute the point estimate of the A-measure, {\em i.e.} the  $\hat{A}$-measure or $\hat{A}_{BC}$, of the two distributions of data samples $B$ and $C$ of sizes $m$ and $n$ respectively. }
    \label{fig:ddr_example_Ameasure}
\end{figure}

In order to estimate the A-measure using samples from two distributions, the point estimate of the A-measure, here denoted the $\hat{A}$-measure (with a hat), is used. (In the Spartan package \cite{Spartan2013}, this is referred to as the A test score).  
For example, if we want to compare two discrete distributions $B$ and $C$, where $B$ comprises $m$ data samples (of some variable $X$) so that $B=\{b_1, b_2, .. , b_m\}$ and $C$ comprises $n$ data samples (of some variable $X$) so that $C=\{c_1, c_2, .. , c_n\}$ then 

\begin{equation}
   \hat{A}_{BC}(X)=\frac{\mathcal{\#}(b_i>c_j)}{mn} + 0.5 \thinspace \frac{\mathcal{\#}(b_i=c_j)}{mn}, 
   \label{eq:sa_peAm_count}
\end{equation}

where $i=1,2,..,m$ and $j=1,2,..,n$ and $\mathcal{\#}(\text{event})$ is the `counting function' that simply denotes the number of times that a certain event occurs when comparing all possible pairs of data samples $(b_i,c_j)$. For clarity, Figure \ref{fig:ddr_example_Ameasure} provides an example of how the $\hat{A}$-measure can be computed by simply counting events. 

Using more conventional mathematical notation, the $\hat{A}$-measure is given by

\begin{equation}
\label{eq:pointestimateA_sum}
   \hat{A}_{BC}(X)=\frac{1}{mn}\mathlarger{\sum}_{i=1}^m  
   \mathlarger{\sum}_{j=1}^n  H(b_i-c_j)  ,
\end{equation}
where $H(x)$ is the Heaviside step function such that

\begin{equation}
    H(x)=\left\{
                \begin{array}{ll}
                  1 \quad \text{   for $x>0$}, \\
                  \frac{1}{2} \quad  \text{   for $x=0$},\\
                  0 \quad  \text{   for $x<0$}.
                \end{array}
              \right.
\end{equation}

If $\hat{A}_{BC}(X)=0.5$, then the distributions $B$ and $C$ are stochastically equal with respect to the variable $X$. 
The $\hat{A}$-measure can thus be used to measure `how equal' two discrete distributions $B$ and $C$ are, by assessing how much the $\hat{A}$-measure $(\in [0,1])$ deviates from equality, {\em i.e.} the value $0.5$. 
The closer the $\hat{A}$-measure is to $0.5$, the `more equal' the two compared distributions are \cite{VarghaDelaney}. 
In many applications, we are only interested in `how equal' two distributions $B$ and $C$ are, and it is not important which distribution is the stochastically greater one. 
In such cases we are only interested in \textit{how much} the $\hat{A}$-measure deviates from stochastic equality ({\em i.e.} the value $0.5$) but the \textit{direction} is not important. Or in mathematical terms: the \textit{magnitude} of the difference between the $\hat{A}$-measure and stochastic equality is important but the \textit{sign} is not. The magnitudal (or scaled) $\hat{A}$-measure (or $\hat{A}$-value), here denoted $\underline{\hat{A}}$ with an underscore, ignores the sign of deviation from equality and is given by 

\begin{equation}
 \underline{\hat{A}} = 
  \begin{cases} 
    \hat{A}_{BC}(X)  & \text{ if $\hat{A}_{BC}(X) \geq 0.5$,} \\
    1-\hat{A}_{BC}(X)& \text{ if $ \hat{A}_{BC}(X) < 0.5$}.
  \end{cases}
  \label{eq:scaled_or_0p5to1}
\end{equation}

\textit{The statistical significance} is used to describe the effect of the stochastic difference between two distributions $B$ and $C$. 
If two distributions $B$ and $C$ are `fairly equal' ({\em i.e.} if they yield an $\underline{\hat{A}}_{BC}$-measure close to 0.5) then the statistical significance is classified as \textit{small}.
The statistical significance is classified using the magnitudal $\underline{\hat{A}}$-measure and, using guidelines from Vargha and Delaney \cite{VarghaDelaney}, the statistical significance is classified to be small, medium or large with respect to $X$ according to the following threshold values for $\underline{\hat{A}}_{BC}(X)$,

\begin{equation}
 \text{Statistical Significance} = 
  \begin{cases} 
    \text{small}  & \text{if $\underline{\hat{A}}_{BC}(X) \in [0.5, 0.56]$}, \\
    \text{medium} & \text{if $\underline{\hat{A}}_{BC}(X) \in (0.56, 0.64]$}, \\
    \text{large} &  \text{if $\underline{\hat{A}}_{BC}(X) \in (0.64, 0.71]$}. 
  \end{cases}
  \label{eq:sa_statsign_cutoffs}
\end{equation}

\begin{figure}
    \centering
    \includegraphics[scale=0.14]{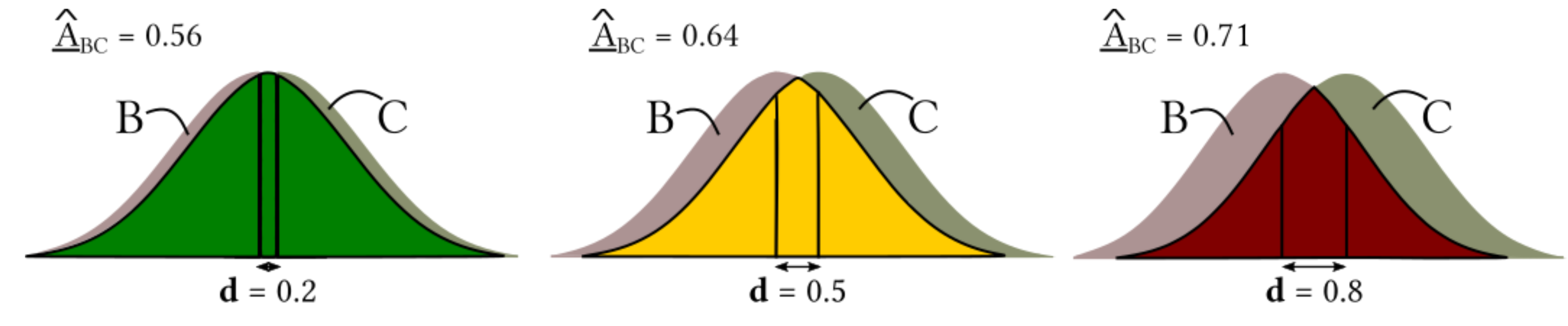}
    \caption{The small (left), medium (centre) and large (right) threshold values for the scaled A-measure of stochastic superiority ($\underline{\hat{A}}_{BC}$) are based on Cohen's \textbf{d}-values comparing two normal distributions $B$ and $C$ with the same variance. The higher the overlap between $B$ and $C$, the smaller the \textbf{d}-value, and the smaller the  $\underline{\hat{A}}_{BC}$-measure ($\underline{\hat{A}}_{BC} \in [0.5,1]$). }
    \label{fig:sa_dvalues}
\end{figure}

These threshold values (that might appear somewhat arbitrary) were first introduced by psychologist and statistician Jacob Cohen \cite{Cohen1962, CohenBook} in the 1960s when comparing normal distributions, but then in terms of another statistical measurement: the effect size (Cohen's) \textbf{d} where

\begin{equation}
    \textbf{d} = \frac{\big| \text{(mean of population $B$) - (mean of population $C$)} \big|}{ \sigma },
\end{equation}

and $\sigma$ is the standard deviation of either $B$ or $C$ (as B and C here are assumed to have the same standard deviation) \cite{CohenBook, RuscioMullen2012}. 
Omitting details from statistics, a small \textbf{d}-value essentially corresponds to a big overlap between distributions $B$ and $C$, whilst a large \textbf{d}-value corresponds to a small overlap between distributions $B$ and $C$, as is illustrated in Figure \ref{fig:sa_dvalues}. Cohen decided to use the threshold \textbf{d}-values for describing `small', `medium' and `large' effect sizes to be 0.2, 0.5 and 0.8 respectively \cite{CohenBook}. 
If we hold on to the assumption that $B$ and $C$ are two normal distributions with the same variability, and furthermore say that they contain the same number of data samples, we can use measures of overlap to get a further `feel' for the previously discussed effect sizes, as illustrated in Figure \ref{fig:sa_dvalues}.  
Cohen's \textbf{d} value can also be converted into \textit{`the probability that a random data sample $X_B$ from (normal) distribution $B$ is larger than a random data sample $X_C$ from (normal) distribution $C$} \cite{McGrawWong}, but that is exactly what the $\hat{A}$-measure $\hat{A}_{BC}(X)$ measures! So this is where the threshold values for the descriptors `small', `medium' and `large' statistical differences, listed in Equation \ref{eq:sa_statsign_cutoffs}, come from. 

\vspace{0.4cm}
Now, Cohen motivated his choice of the \textbf{d}-value thresholds using a blend of intuitive `everyday' examples and mathematical reasoning \cite{CohenBook}. However, he did issue a warning regarding the fact that the threshold values should be determined based on the research methodology at hand. 
Thus the modeller should not blindly use Cohen's suggested thresholds, but instead reason what constitutes a small enough statistical significance in the study at hand. The modeller must also decide how fine the data samples in the data distributions should be before performing Consistency Analysis. In many applications, it is likely that the amount of data samples required in order to achieve a small statistical significance increases with the fineness of the data.
Nonetheless, scientific conventions are useful and thus in the remainder of this review we will use the threshold values suggested by Cohen, as is done in other mathematical biology studies \cite{Spartan2013}.

\section{Consistency Analysis }
\label{sec:sa_ca}
{\it In silico} simulations based on mathematical models with built-in stochasticity will not produce the same output data every simulation run. 
Consistency Analysis (also called aleatory analysis) is a stochastic technique that answers the question: how many data samples do we need to produce in order to mitigate uncertainty originating from intrinsic model stochasticity? In our case, \textit{one data sample} is the product of \textit{one in silico simulation}, so an equivalent question is: how many {\it in silico} simulations should we run before describing our results in terms of, for example, average values, standard deviations or similar? 
\\

Let us say that one \textit{in silico} simulation produces one data sample of some output response $X$. This data sample can for example correspond to `the population size at time point $T$', or something similar. It is up to the modeller to identify and decide what the meaningful output response(s) should be, and Consistency Analysis can be performed on multiple output responses at multiple time steps, for comprehensiveness. Before we begin, note that when performing Consistency Analysis, we always use the calibrated model parameters. 
\\

The first step involved in performing Consistency Analysis is to produce multiple distributions of data of various sizes. We say that a distribution with $n$ data samples has a distribution size $n$, and the goal of Consistency Analysis is to find the smallest $n$ value (here denoted $n^*$) that yields a small stochastic significance. To do this, we create various \textit{distribution groups} that all contain 20 distributions each of some distribution size $n$, as is shown in \textbf{Step 1} in Section \ref{sec:sa_ca_lazy}. Following the methodology described in previous work by Alden {\em et al.}, and the Spartan package that they developed \cite{Spartan2013}, we create one distribution group that contains 20 distributions of size $n=1$, one distribution group that contains 20 distributions of size $n=5$ and so on.  Here, the $n$ values 1, 5, 50, 100 and 300 are evaluated \cite{Spartan2013} and thus we must produce a total of $20 \times (1+5+50+100+300)=9120$ {\it in silico} runs. (Note that if the desired accuracy is not achieved for the highest investigated $n$ value, here $n=300$, higher values of $n$ can be explored). 
\\

We here let a distribution $D_{n,k}$ denote the $k$th distribution of distribution size $n$ so that 

\begin{equation}
D_{n,k}=\{d_{n,k}^1, d_{n,k}^2, \: .. \: ,d_{n,k}^n \}
\end{equation}

where $d_{n,k}^h$ is the the $h$th data sample in distribution $D_{n,k}$ and $h=1,2,..,n$. 
The $\hat{A}$-measure resulting from comparing two distributions $D_{n,k}$ and $D_{n,k'}$ with respect to the variable $X$ is denoted by $\hat{A}^{n}_{k,k'}(X).$ 
\\

Now, within every distribution-group, we compare the first distribution ($k=1$) to all other distributions ($k'=2,3,..,20$) using the $\hat{A}$-measure. 
This yields 19 $\hat{A}$-measures per distribution-group, as is shown in \textbf{Step 2} in Section \ref{sec:sa_ca_lazy}. The maximum scaled $\hat{A}$-measure with respect to $X$, occurring in a distribution-group $g$ that contains distributions of size $n_g$, is denoted $\underline{\hat{A}}^{n_g}_{max}(X)$. The smallest value $n_g$ for which $\underline{\hat{A}}^{n_g}_{max}(X) \leq 0.56$ is denoted $n^*$. 
In other words: $n^*$ corresponds to the smallest distribution size for which all of the 19 computed $\hat{A}$-measures yield a small stochastic significance, as is shown in \textbf{Step 3} in Section \ref{sec:sa_ca_lazy}.  
This answers the question that we set out to answer via Consistency Analysis: 
$n^*$ data samples (or {\it in silico} runs) are needed in order to mitigate uncertainty originating from intrinsic model stochasticity.
The procedure on how to perform Consistency Analysis is outlined Section \ref{sec:sa_ca_lazy}.

\subsection{Quick Guide: Consistency Analysis}
\label{sec:sa_ca_lazy}
Here follows a quick guide for how to perform Consistency Analysis.

\vspace{0.5cm}

\includegraphics[scale=0.37]{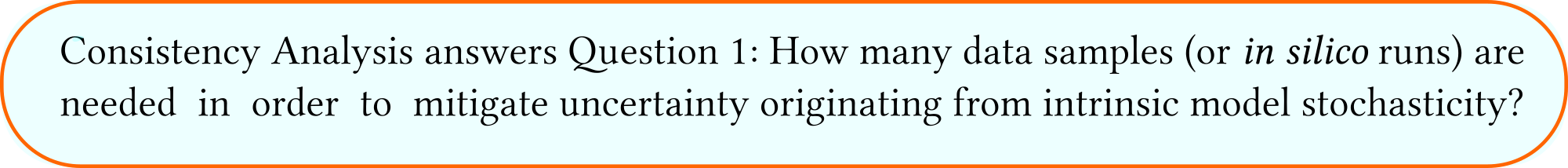}
\vspace{0.5cm}

\includegraphics[scale=0.37]{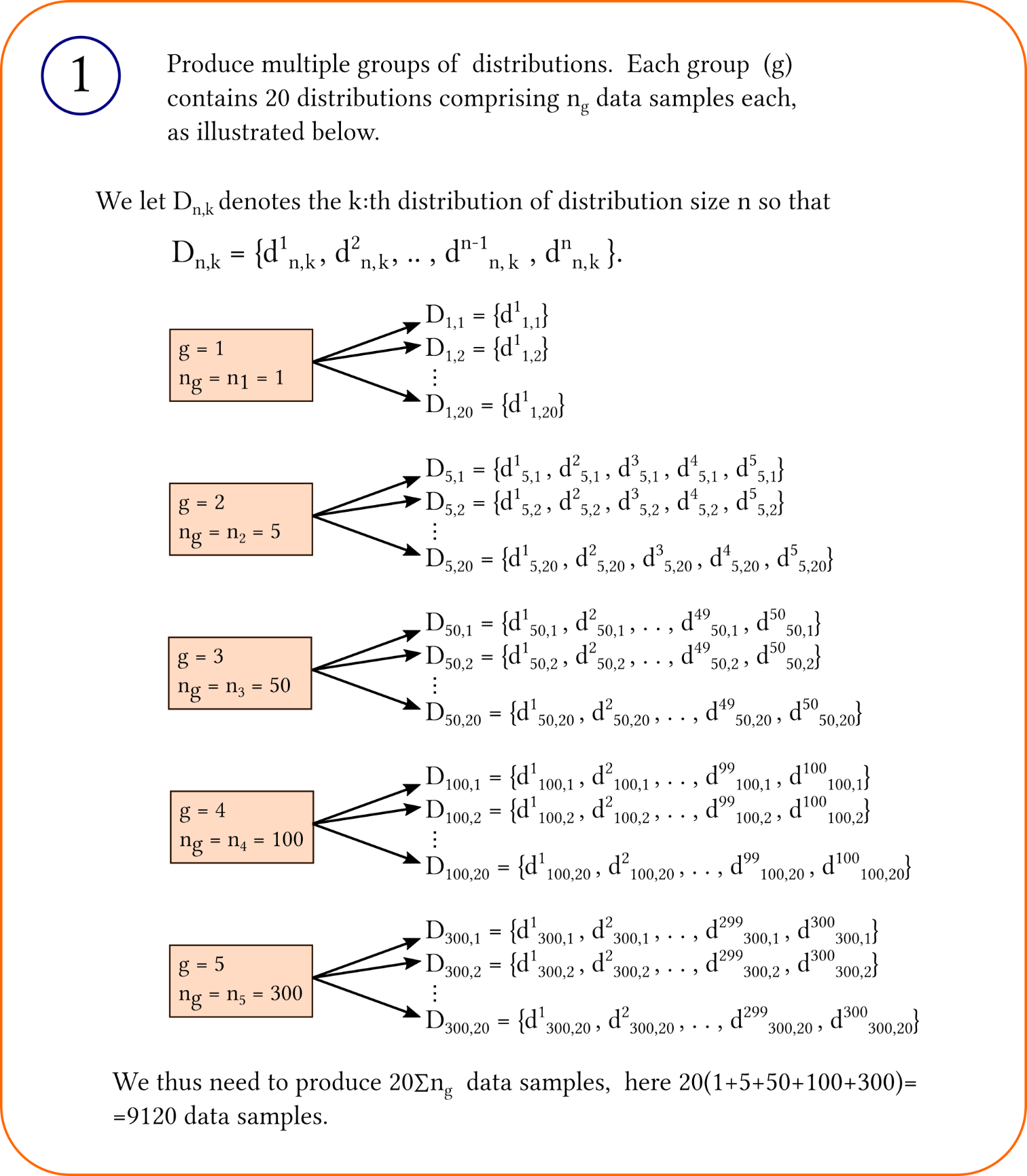}

\includegraphics[scale=0.74]{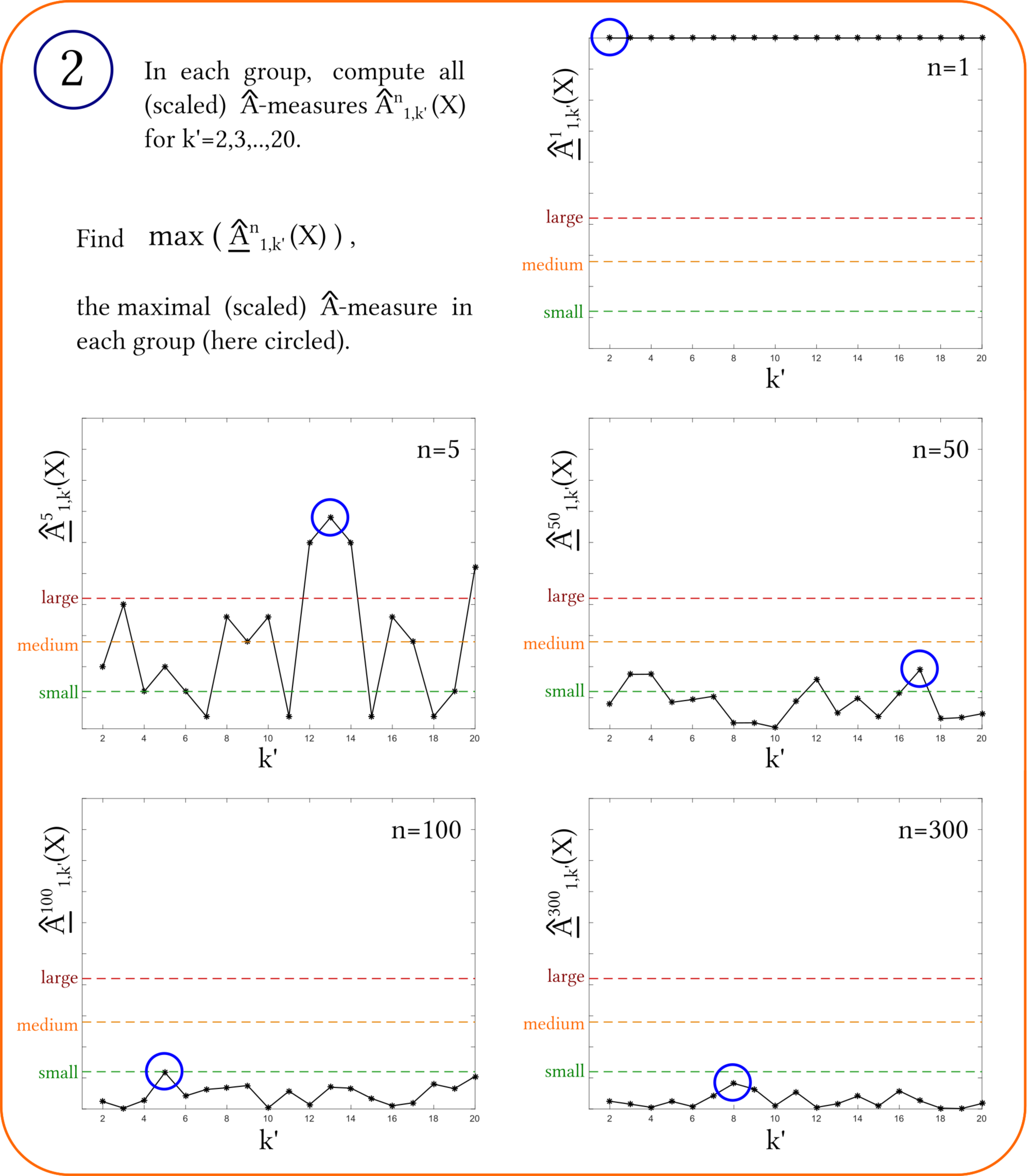}

\includegraphics[scale=0.37]{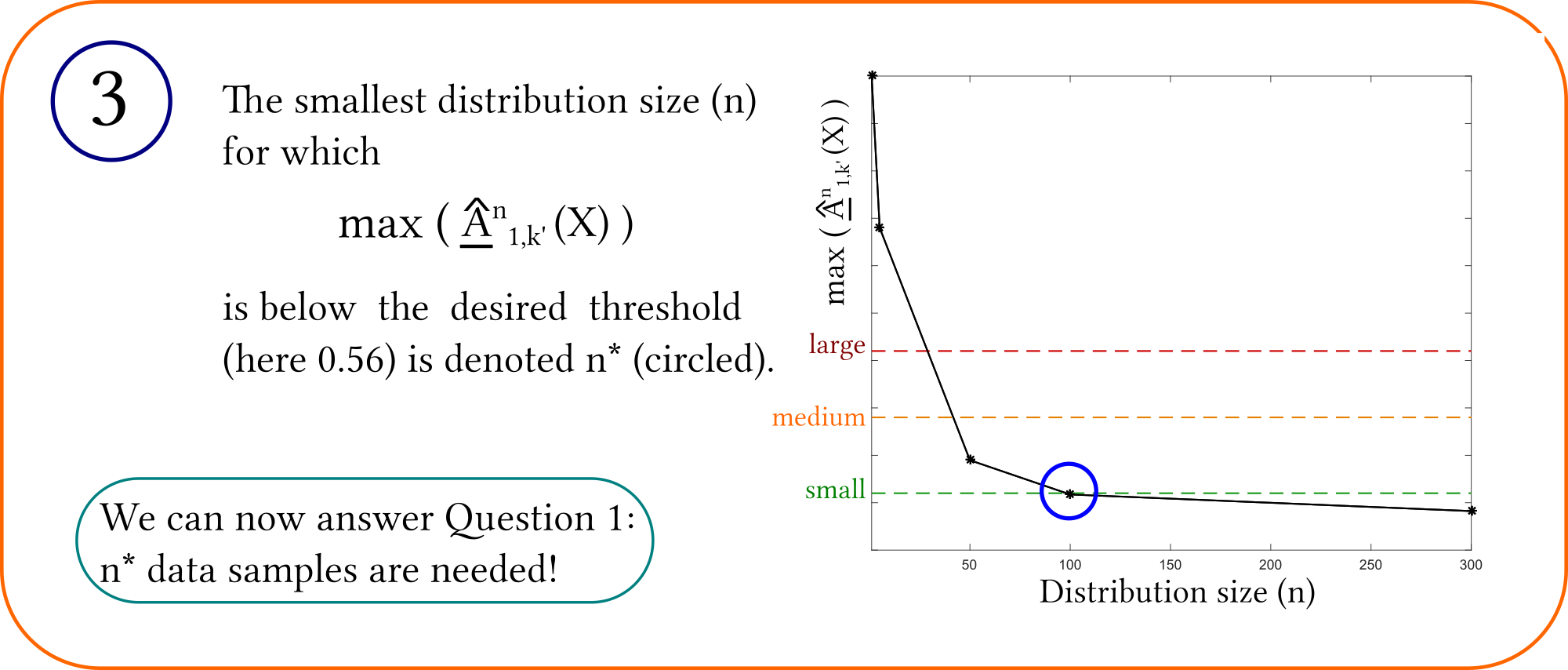}

\vspace{0.5cm}

\includegraphics[scale=0.37]{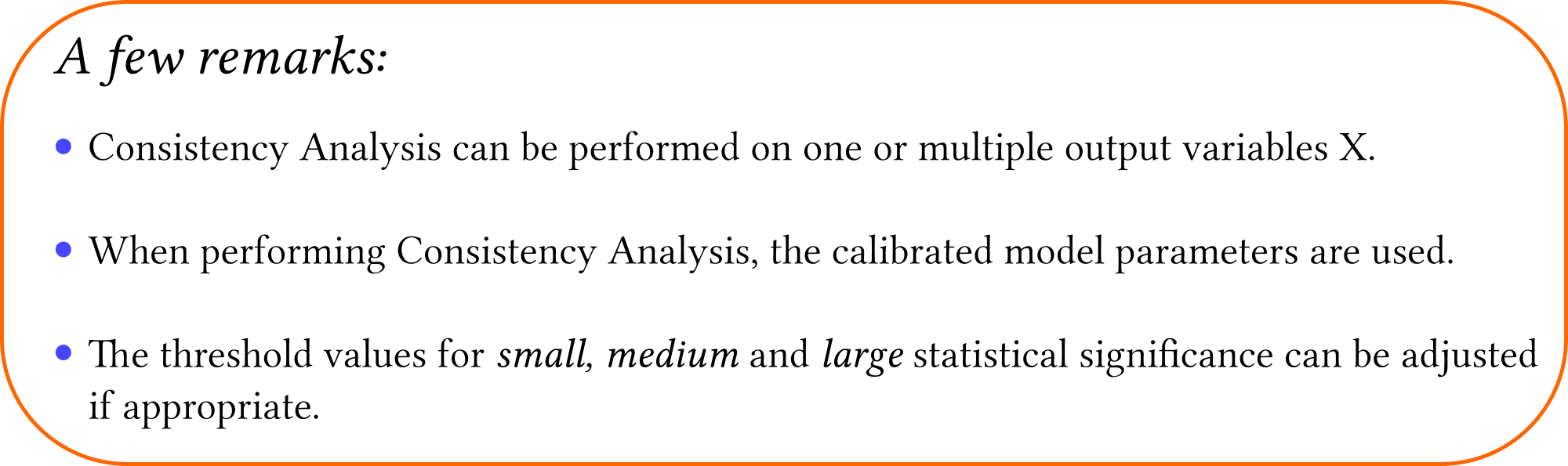}

\section{Robustness Analysis}
\label{sec:sa_ra}
Robustness Analysis answers the question: how robust are model responses to {\it local} parameter perturbations? 
Robustness Analysis investigates if, and how, perturbing the value of \textit{one} input parameter significantly changes an output $X$. 
								%
Using the ${\hat{A}}$-measure, data distributions containing output data produced by perturbed input parameters, are compared to a data distribution containing output data produced by the calibrated input parameters.  
All perturbed data distributions are here of size $n^*$, where $n^*$ is decided in the Consistency Analysis process, previously described in Section \ref{sec:sa_ca}, when analysing stochastic models. 
%
\\

Before commencing the Robustness Analysis, we must identify the uncertain model parameters that we want to investigate the robustness of. 
We denote these parameters $p^i$, where $i=1,2,..,q$, and thus we have a total of $q$ parameters whose robustness we will investigate.  
Now, as illustrated in \textbf{Step 1} in Section \ref{sec:sa_qg_ra}, we let each such parameter $p^i$ be investigated at $r(p^i)$ different parameter values (including the calibrated value), and thus we need to generate a total of $P$ distributions of sample size $n^*$ where

\begin{equation}
P=\sum_{i=1}^{q} r(p^i).
\end{equation}
								
Note that the number of investigated parameter values, $r(p^i)$, need not be the same for every input parameter $p^i$. 
Investigated distributions of sample size $n^*$ are here denoted $D_{n^*,p^i_j}$, where $i=1,2,..,q$ denotes which parameter is being perturbed and $j=1,2,..,r(p^i)$ denotes the specific perturbation of parameter $p^i$. 
For some perturbation $j=C$, the parameter value $p^i_j$ equals the calibrated value for input parameter $p^i$. 
For each parameter that we are investigating, the $\hat{A}$-measure is used to compare the calibrated distribution $D_{n^*,p^i_C}$ to all distributions $D_{n^*,p^i_j}$. Note that when $j=C$, the calibrated distribution is compared to itself and thus the $\hat{A}$-measure equals 0.5. 
These $\hat{A}$-measures provide information regarding the statistical significance, specifically if it can be described to be \textit{small, medium} or \textit{large} under parameter perturbations. 
Plotting the corresponding $\hat{A}$-measure over the parameter value $p^i_j$ for each parameter $p^i$, paints an informative picture of local parameter robustness, as shown in \textbf{Step 2}, in Section \ref{sec:sa_qg_ra}.
Another descriptive way to demonstrate the influence that parameter values $p^i_j$ have on some output response $X$ is to use boxplots. 
As is illustrated in \textbf{Step 3} in Section \ref{sec:sa_qg_ra}, boxplots can be used to clearly show the median, different percentiles, and outliers of some data distribution $D_{n^*,p^i_j}$ as a function of the parameter value $p^i_j$. 
The methodology to perform Robustness Analysis is outlined in Section \ref{sec:sa_qg_ra}.  
Note that Robustness Analysis does not pick up on any non-linear effects between an input parameter $p^i$ and an output $X$, that occur when more than one model parameter is simultaneously perturbed \cite{Charzynska2012}. 
Such effects can however be identified using a global sensitivity analysis technique, such as Latin Hypercube Analysis, as described in Section \ref{sec:sa_lhc}. 

\subsection{Quick Guide: Robustness Analysis}
\label{sec:sa_qg_ra}

\includegraphics[scale=0.4]{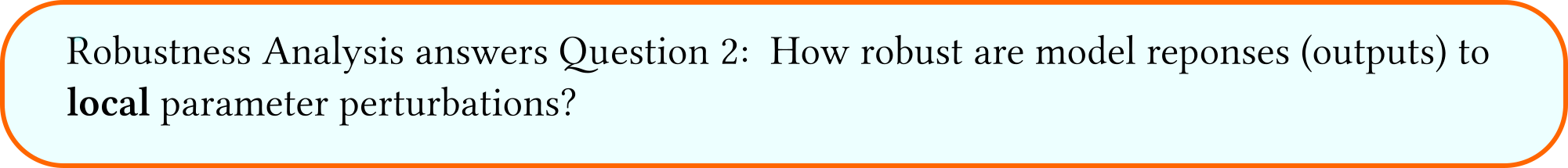}
\\

\includegraphics[scale=0.8]{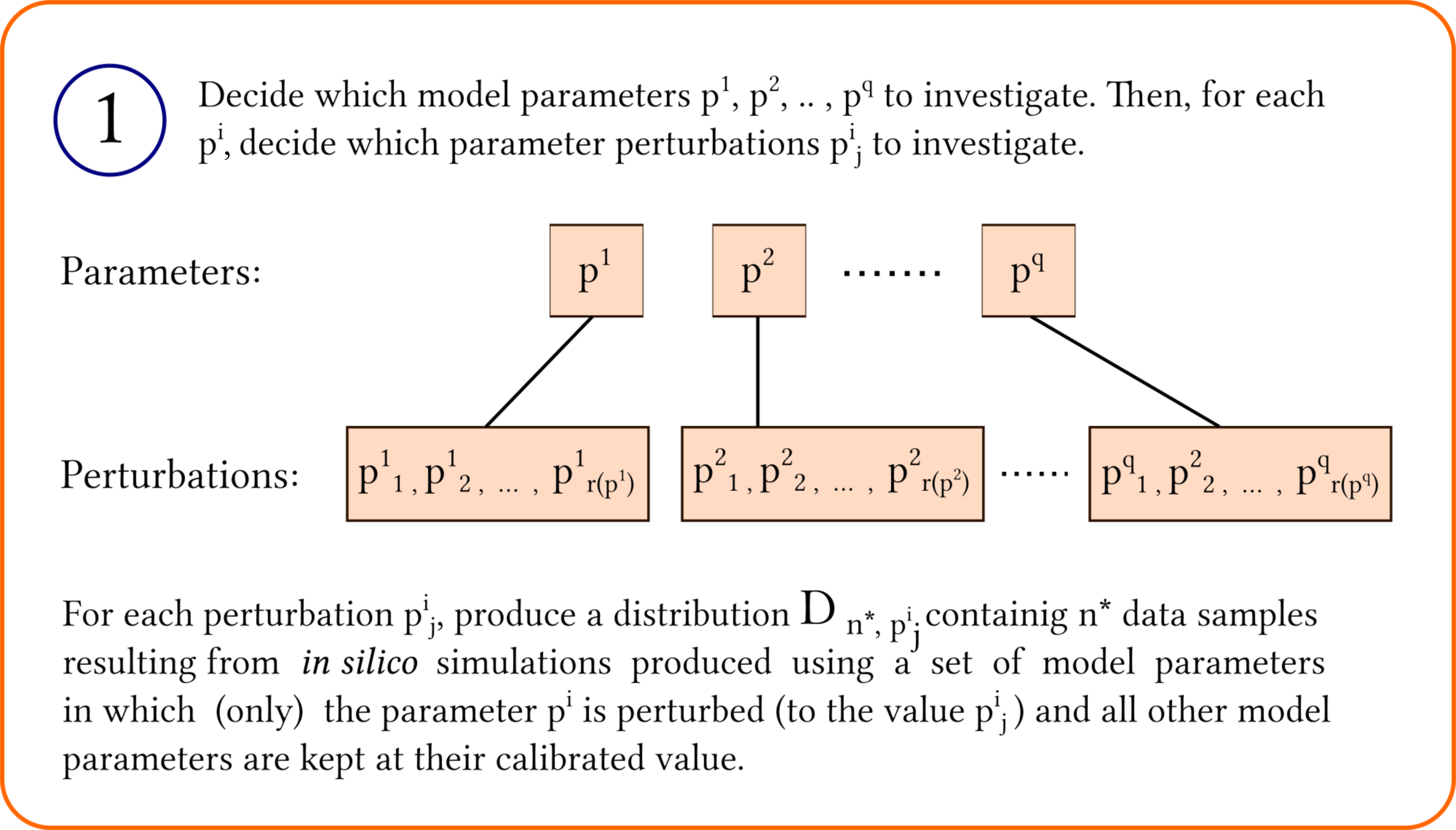}
\\

\includegraphics[scale=0.4]{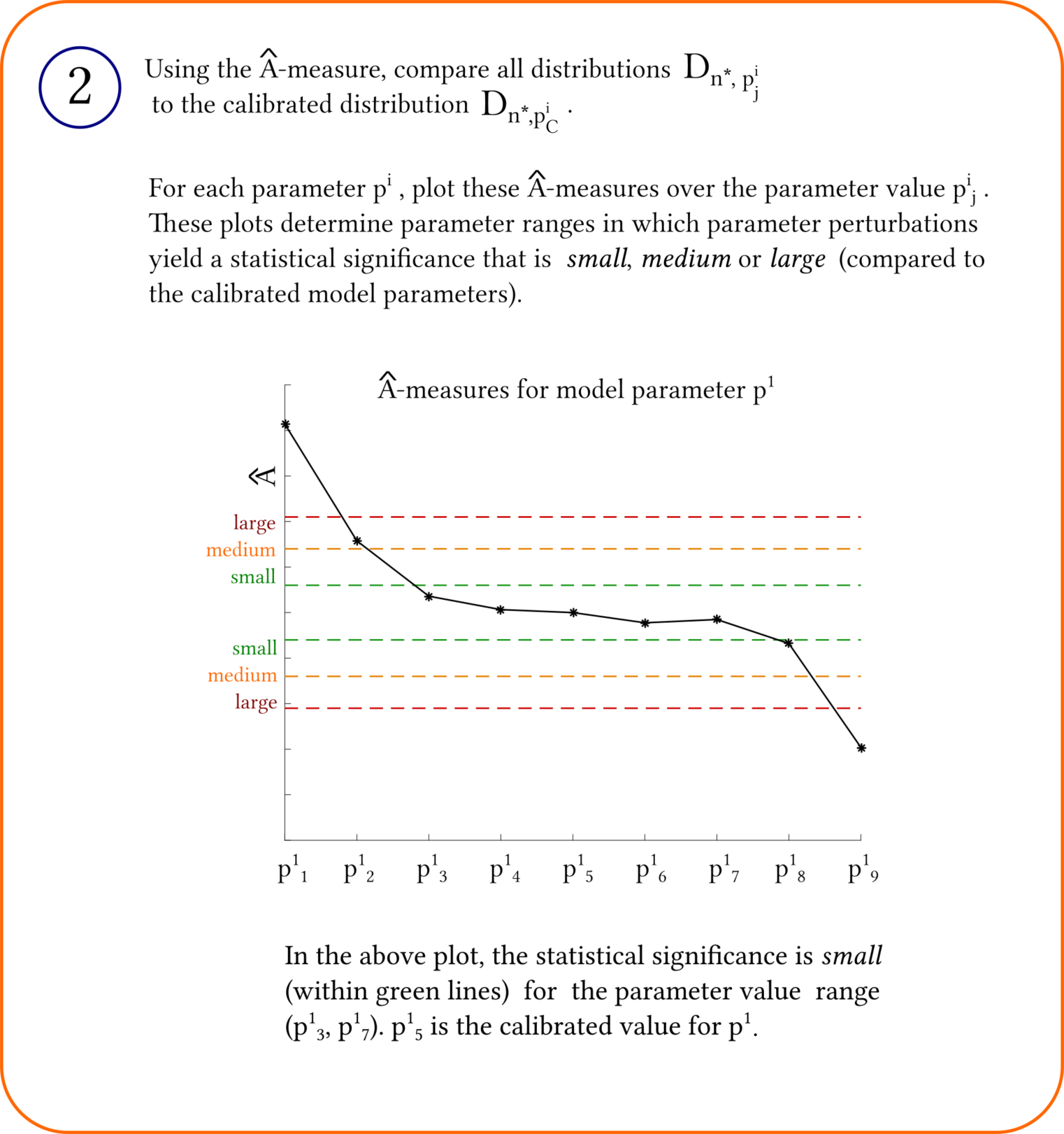}
\\

\includegraphics[scale=0.8]{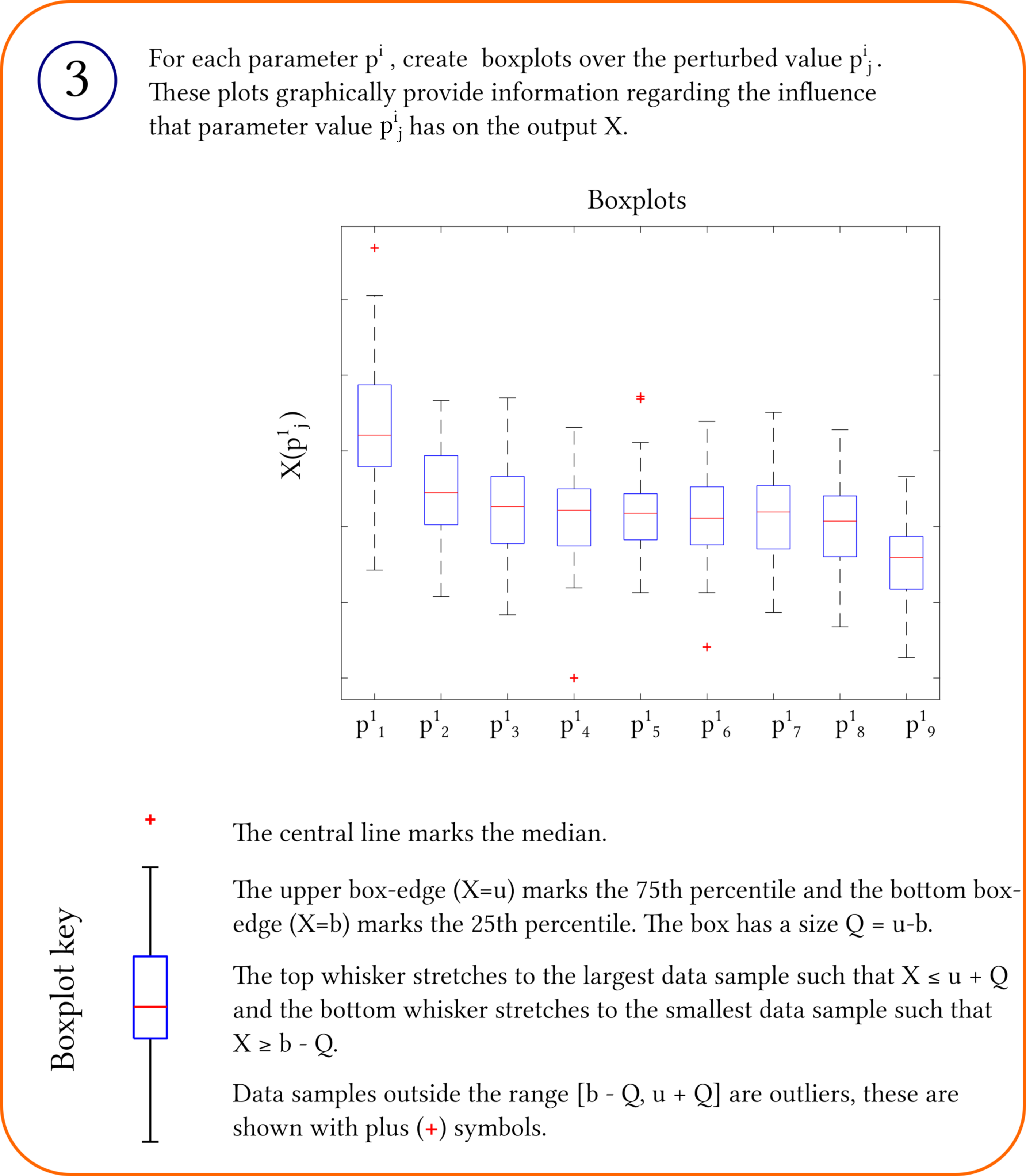}
\\

\includegraphics[scale=0.4]{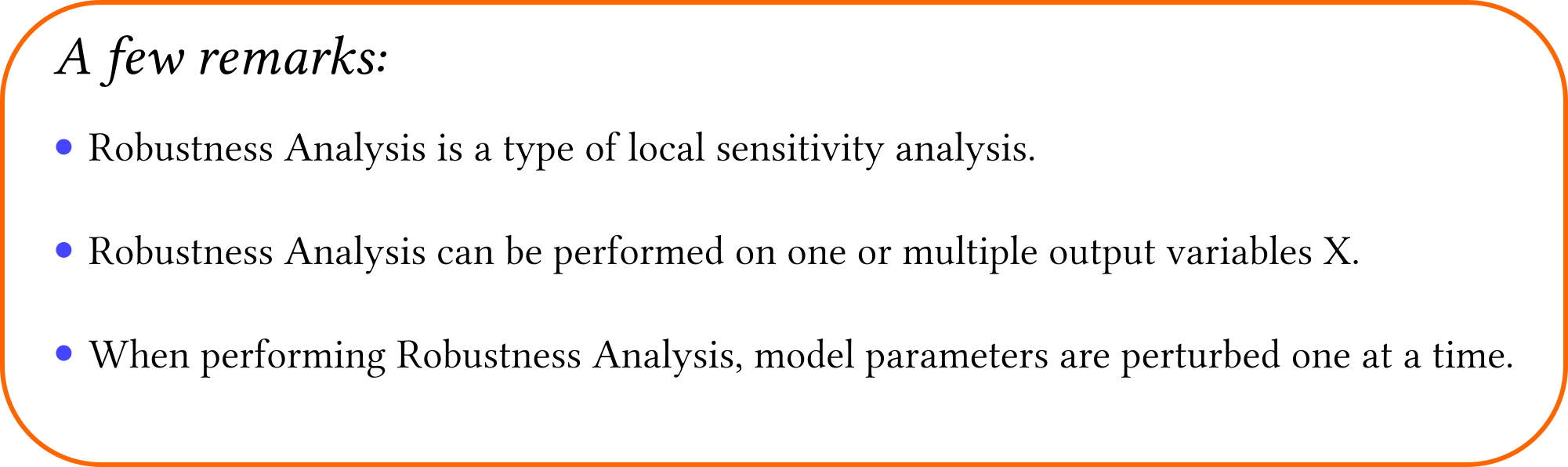}

    
    
    
    

\section{Latin Hypercube Sampling and Analysis}
\label{sec:sa_lhc}
Latin Hypercube Analysis answers the question: how robust are model responses to {\it global} parameter perturbations?
Latin Hypercube Analysis is a type of global sensitivity analysis that investigates the relationship between input parameters and output responses when all input parameters are simultaneously perturbed. 
The parameters that we want to perturb are (as in Section \ref{sec:sa_ra}) denoted $p^i$, where $i=1,2,..,q$. 
Thus the parameters $p^1, p^2, ... , p^q$ together span a parameter space of dimension $q$.
%
It is impossible to test every possible combination of input parameter values if they are picked from continuous ranges. In fact, even if we select a finite number of parameter values $r(p^i)$ to test for each parameter $p^i$, or if we pick discrete parameter values, comparing every possible combination of parameter values may require us to produce an impractically large number of simulation runs. Thus performing {\it in silico} simulations for all possible combinations of input parameters will in many cases be at worst impossible, and at best impractical. 
In order to circumvent this issue, Latin Hypercube Sampling can be used \cite{Spartan2013}. It is a sampling technique that ensures comprehensive testing coverage over the parameter space whilst keeping the number of tested parameter combinations low enough to be applicable in practice \cite{McKay1979, McKay1992}.  
After Latin Hypercube \textit{Sampling} (Section \ref{sec:sa_lhc_s}), Latin Hypercube \textit{Analysis} (Section \ref{sec:sa_lhc_a}) is used in order to assess global sensitivity.

\subsection{Latin Hypercube Sampling}
\label{sec:sa_lhc_s}
\begin{figure}[H]
    \centering
    \includegraphics[scale=0.275]{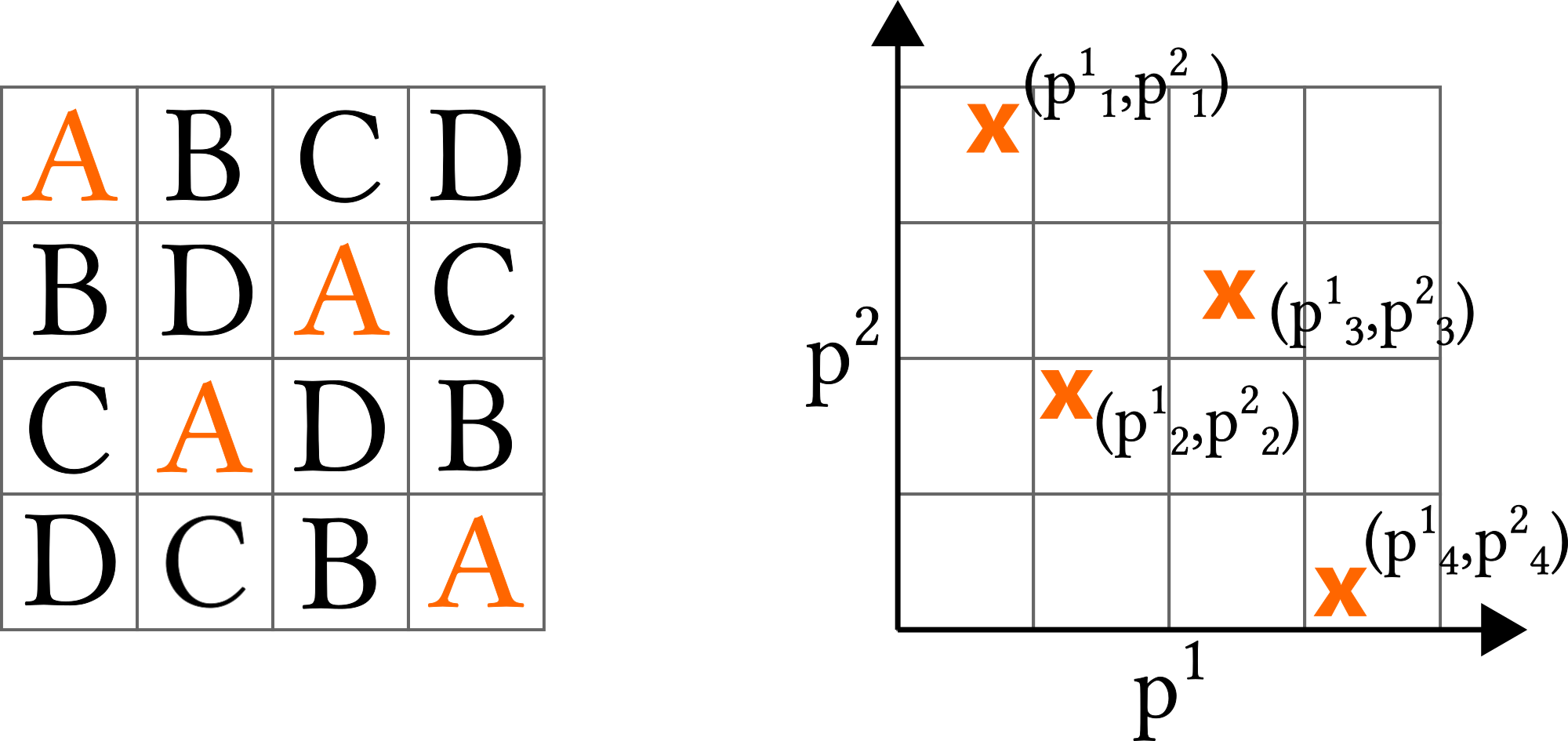}
    \caption{Left: An $\ell \times \ell$ Latin Square in which each Latin symbol occurs $\ell$ times, exactly once in each row and exactly once in each column. Right (analogously): A two-dimensional parameter space spanned by the input parameters $p^1$ and $p^2$ that are both sectioned into $\ell$ intervals. Using Latin Hypercube sampling, $\ell$ parameter combinations $(p^1_j, p^2_j)$ are sampled where $j=1,2,..,\ell$ and each $p^1$-interval is sampled from exactly once and each $p^2$-interval is sampled from exactly once.  }
    \label{fig:sa_lhc_square}.
\end{figure}

In the two-dimensional case, a \textit{Latin Square} is an $\ell \times \ell$ square grid containing $\ell$ (traditionally Latin, hence the name) different symbols such that each symbol occurs exactly once in every row and exactly once in every column \cite{Sheikholeslami2017}, as illustrated in Figure \ref{fig:sa_lhc_square}. Analogously, in the Latin Hypercube Sampling framework, consider two parameters $p^1$ and $p^2$, spanning a parameter space of dimension $q=2$, where both $p^1$ and $p^2$ are sectioned into $\ell$ intervals. %
We then pick $\ell$ combinations of input parameter values (or sampling points) $(p^1_j, p^2_j)$, where $j=1,2,...,\ell$, such that every $p^1$-interval is sampled from exactly once and every $p^2$-interval is sampled from exactly once. Within the parameter range of an interval, the sampled parameter value $p^i_j$ is randomly selected (unless of course the interval contains only one possible value $p^i_j$). 
Note that the $j$ index denotes the coordinate combination that $p^i_j$ belongs to, not the interval from which the parameter value $p^i_j$ was taken. Thus there is no condition demanding that the values $p^i_j$ are ordered in a way such that $p^i_1 < p^i_2 < ... < p^i_\ell$.

The analogy between a Latin Square and Latin Hypercube Sampling from a two-dimensional parameter space is illustrated in Figure \ref{fig:sa_lhc_square}. The Latin Square can be extended to higher dimensions to form a Latin Cube (dimension = 3) or a Latin Hypercube (dimension > 3) and, analogously, the two-dimensional sampling space illustrated in Figure \ref{fig:sa_lhc_square} can be extended to $q$ dimensions, spanned by the input parameters $p^1, p^2, .. ,p^q$ \cite{Sheikholeslami2017}. 

For each parameter $p^i$, the total investigated parameter range is $[min(p^i), \; max(p^i)]$, where $min(p^i)$ and $max(p^i)$ respectively denote the minimum and maximum values of $p^i$ to be investigated. Now each parameter range $[min(p^i), \; max(p^i)]$ is sectioned into $N$ intervals, and we denote these intervals by $u_{p^i}^1, u_{p^i}^2, ....,u_{p^i}^N$. Note that all input parameters $p^i$ must be sectioned into the same number of intervals. 
If the intervals are of equal size, then the size of an interval, $w(p^i)$, is 

\begin{equation}
    w(p^i)=\frac{max(p^i)-min(p^i)}{N}
\end{equation}

and the $r$th interval $u_{p^i}^r$ has a parameter range such that

\begin{equation}
    u_{p^i}^r= \big[min(p^i)+w \cdot (r-1), \; min(p^i)+w \cdot r  \big]
\end{equation}

where $r=1,2,...,N$.
\\

Note that there are more than one way to populate Latin symbols in a Latin Square, this can be realised by regarding Figure \ref{fig:sa_lhc_square} and noticing that the \textbf{A}-symbols and the \textbf{B}-symbols cover the Latin Square in different ways. 
Analogously, and by extension, there are multiples ways to populate sampling coordinates in a Latin Hypercube Sampling framework. Some of these ways provide better coverage of the parameter space than do others \cite{Sheikholeslami2017}, but details regarding such sampling-optimisation are outside the scope of this review. 
Here, we use the built-in \textsf{MATLAB} function \textsf{lhsdesign} \cite{MATLAB:2019b} to select which parameter combinations to use according to a Latin Hypercube Sampling approach, details about the implementation are available in the Appendix. 
Note that, in our case, all $N$ intervals $u_{p^i}^1, u_{p^i}^2, ....,u_{p^i}^N$ for a parameter $p^i$ are uniformly spaced, but the choice of spacing can be adjusted to the specific application at hand \cite{MATLAB:2019b}. 
%
\\

Now let us address the choice of intervals $N$, as this is not straightforward. 
Using the Latin Hypercube Sampling framework, every parameter $p^i$, where $i=1,2,..,q$, is partitioned into $N$ intervals and, consequently, $N$ combinations comprising $q$ parameter values are sampled and tested. 
Compared to a small $N$ value, a large value of $N$ will provide more data to use, and draw conclusions from, in the Latin Hypercube \textit{Analysis} stage, however, it will also increase the computational cost in the  Latin Hypercube \textit{Sampling} stage. 
There is no strict rule for how to choose $N$, but suggested values for $N$ in the literature are $N=2q$ for large values of $q$ ({\em i.e.} high-dimensional parameter spaces) 
or $N=4q/3$ which has been described to be `usually satisfactory'  \cite{Manache2007, imanhelton1985}. 
Authors of the Spartan package use a lot larger numbers in their provided examples \cite{Spartan2013}.
In this example study, we decide to use $N=100$ uniform intervals. 
At the end of the day, the choice of $N$ is up to the modeller, who must outweigh the (computational) cost of producing a large number of data samples, with the advantage of having a vast amount of data, and thus plentiful information, in the analysis stage. Details regarding quantitative choices of $N$ are outside the scope of this review. 
%


\subsection{Latin Hypercube Analysis}
\label{sec:sa_lhc_a}
During the Latin Hypercube Sampling process, $N$ different points in the $q$-dimensional parameter space spanned by the input parameters $p^1,p^2,...,p^q$ are selected as \textit{sampling points}, as shown in \textbf{Step 1} in Section \ref{sec:sa_lhc_lazy}. One such sampling point, $C_j$, can be described by its coordinates in the parameter space so that $C_j=(p^1_j, p^2_j, ...,p^q_j)$. Each sampling point $C_j$ is used to generate $n^*$ output responses $X(C_j)$, where $n^*$ is determined using Consistency Analysis. Subsequently, the median output value, here denoted $\underaccent{\tilde}{X}(C_{j})$, is computed for every $C_{j}$. 
Now, our overall aim is to investigate the relationship between an input parameter $p^i$ and an output response ${X}$.
We investigate this input-output relationship in two steps, one of which is qualitative and one of which is quantitative. 
In the first and qualitative step, we produce two-dimensional scatterplots in which median output data, 
$$\underaccent{\tilde}{X}(C_{1}), \underaccent{\tilde}{X}(C_{2}),
...,\underaccent{\tilde}{X}(C_{N})
= 
\underaccent{\tilde}{X}(p^1_1, p^2_1, ..., p^q_1), \underaccent{\tilde}{X}(p^1_2, p^2_2, ..., p^q_2),
...,\underaccent{\tilde}{X}(p^1_N, p^2_N, ..., p^q_N), $$

are plotted over parameter values 

$$p^i_1, p^i_2, ..., p^i_N,$$ 

for one of the input parameters $p^i$. We do this for every input parameter $i=1,2,...,q$ and thus $q$ scatterplots are created. 
By simply visually analysing the data in the scatterplots, we are able to make qualitative observations regarding the relationship between the input and the output. Examples of such observations are provided in \textbf{Step 2} in Section \ref{sec:sa_lhc_lazy}
\\

As a second step, we use a quantitative measure, such as the Pearson Product Moment Correlation Coefficient (or the correlation coefficient for short), to quantitatively describe the correlation between input parameters and output responses, as done in \textbf{Step 3} in Section \ref{sec:sa_lhc_lazy}. 
The correlation coefficient is denoted $r$, where $r \in [-1,+1]$. It describes the linear association between the input parameter and the output response in terms of both magnitude and direction. 
A positive (linear) correlation between $p^i$ and $\underaccent{\tilde}{X}(C_j)$ means that if either the input value or the output value increases, so does the other one, and thus $r$ is positive. 
Conversely, a negative correlation means that if either $p^i$ or $\underaccent{\tilde}{X}(C_j)$ increases, the other one decreases, and thus $r$ is negative. The magnitude of $r$ describes the strength of the correlation, where a magnitude of $1$ corresponds to a strong linear association, and a small magnitude corresponds to a weak correlation. An $r$ value of approximately zero indicates that there is no linear correlation between the two investigated variables. Note that the Pearson Product Moment Correlation Coefficient picks up linear associations only, thus there may exist other, non-linear correlations that are not captured by the correlation coefficient $r$. 
Therefore it is important to, not only quantitatively compute input-output correlations, but to also qualitatively assess the relationships between inputs and outputs via data visualisation in scatterplots as previously described.
\\

The correlation coefficient, $r^i$, describing the correlation between an input parameter $p^i$, and an output response $X$ (in median form) is given by \cite{Mukaka2012},

\begin{equation}
    r^i=\frac{\mathlarger{\mathlarger{\mathlarger{\sum}}}_{j=1}^N \bigg( p^i_j - \bar{p}^i \bigg)
    \bigg(\underaccent{\tilde}{X}(C_j) - \underaccent{\tilde}{\bar{X}} \bigg) } 
    {\sqrt{\bigg{(} \mathlarger{\mathlarger{\mathlarger{\sum}}}_{j=1}^N \big( p^i_j - \bar{p}^i \big)^2 \bigg{)}
    \bigg{(} \mathlarger{\mathlarger{\mathlarger{\sum}}}_{j=1}^N
    \big( \underaccent{\tilde}{X}(C_j) - \underaccent{\tilde}{\bar{X}} \big) ^2
    \bigg{)}
    }
    },
\end{equation}
where a bar denotes the mean value.
\newpage



%
%
%
When it comes to interpreting quantitative input-output relationships based on the correlation coefficient $r$, there are no all-encompassing threshold values to use for descriptors such as `weak', `moderate', `strong' \cite{Mukaka2012, Schober2018, Krehbiel2004}. Relationships quantified by correlation coefficient values close to the extrema 0 or 1 may be easy to describe as `negligible' or `strong', respectively. However, correlation coefficient values in the middle of the [0,1] range are more difficult to label. Various `rules of thumb' have been suggested in the literature but, at the end of the day, it is up to the modeller to appropriately judge what constitutes a `weak', `moderate' or `strong' input-output relationship in the specific (modelling) application at hand, taking into account the research area, the number of data samples, and the range of investigated input values \cite{Schober2018}. However, even without rigid descriptor threshold values, we can compare the correlation coefficient values for all input-output pairs and see which input values are the most influential within the ranges of regarded input values. As a guide, suggested correlation coefficient descriptor threshold values presented in the literature are listed in Table \ref{tab:sa_corrcoeff_thresh}.  
The methodology to perform Latin Hypercube Sampling and Analysis is outlined in Section \ref{sec:sa_lhc_lazy}.
\\

\begin{table}[H]
\label{tab:sa_corrcoeff_thresh}
\begin{center}
\setlength{\extrarowheight}{3pt}
\begin{tabular}{|x{3.4cm}||x{1.9cm}|x{1.9cm}|x{1.9cm}|x{1.9cm}|x{1.9cm}|}
\hline
       \diaghead(-3,2){\hskip \hsize}{reference}{descriptor}& negligible & weak & moderate & strong & very strong \\ \hline
    Mukaka \cite{Mukaka2012}& [0,0.3) & [0.3,0.5) & [0.5,0.7) & [0.7,0.9) & [0.9,1]\\ \hline
    Schober {\em et al.} \cite{Schober2018} & [0,0.1) & [0.1,0.4) & [0.4,0.7) & [0.7,0.9) & [0.9,1] \\ \hline
    Krehbiel \cite{Krehbiel2004} & \multicolumn{5}{|l|}{``A linear relationship exists if $|r|\geq 2 / \sqrt{\text{number of samples}}$.''} \\ \hline
  
\hline
\end{tabular}
\end{center}

\label{tab:ca}
\caption{Suggested descriptor threshold values for the magnitude of the correlation coefficient, $|r|$, reported in the literature.  }
\end{table}
\normalsize

\subsection{Quick Guide: Latin Hypercube Sampling and Analysis}
\label{sec:sa_lhc_lazy}

\includegraphics[scale=0.4]{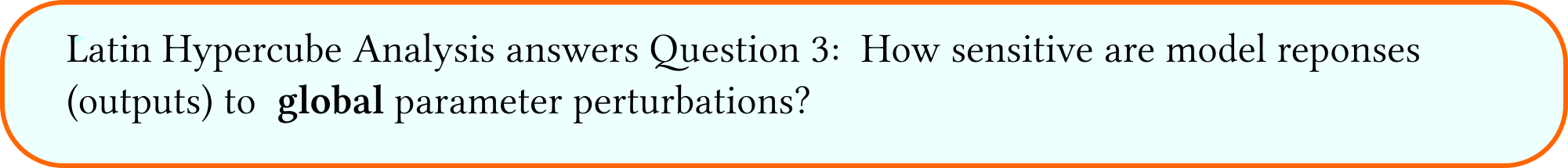}
\\

\includegraphics[scale=0.8]{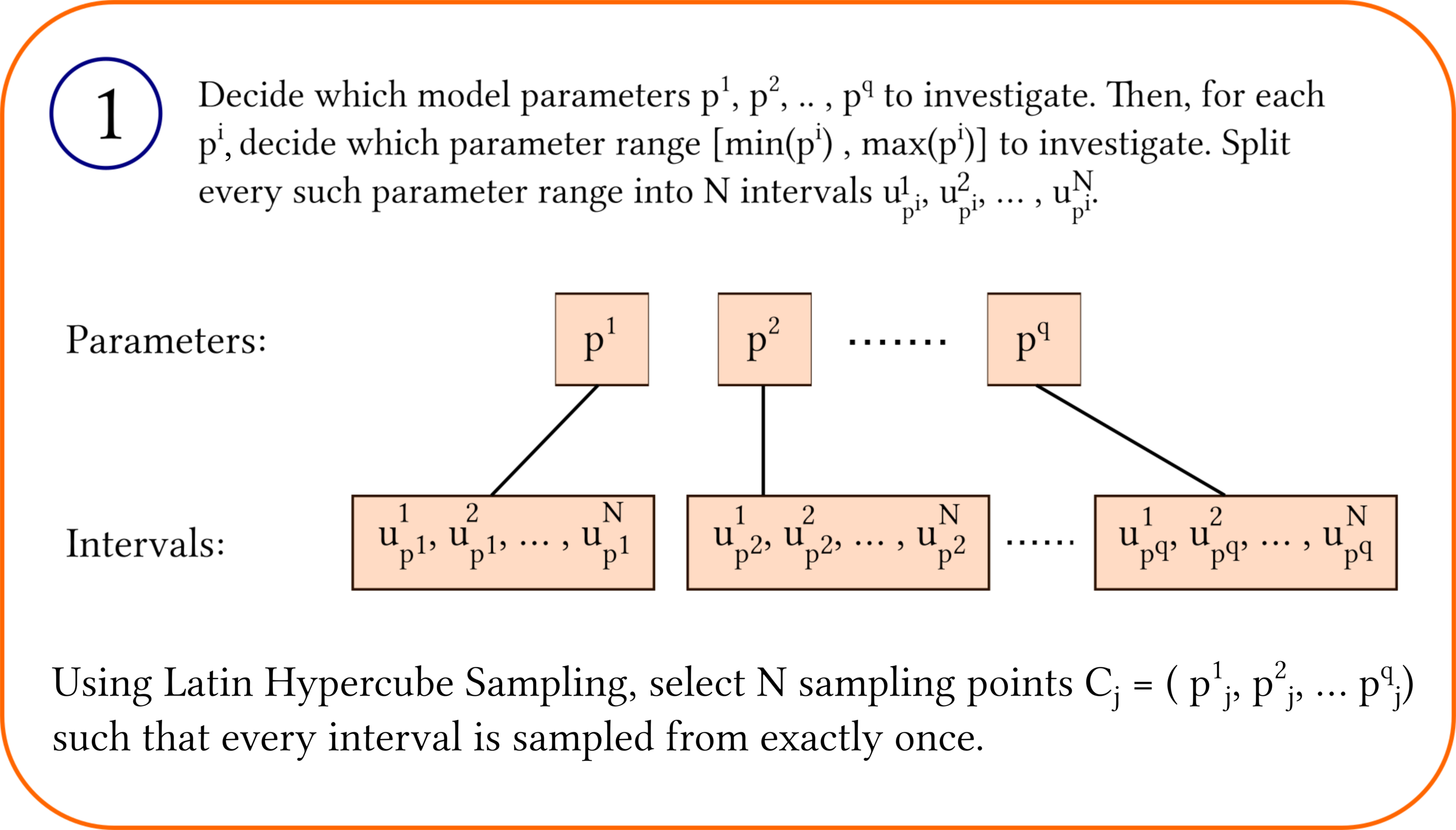}
\\

\includegraphics[scale=0.4]{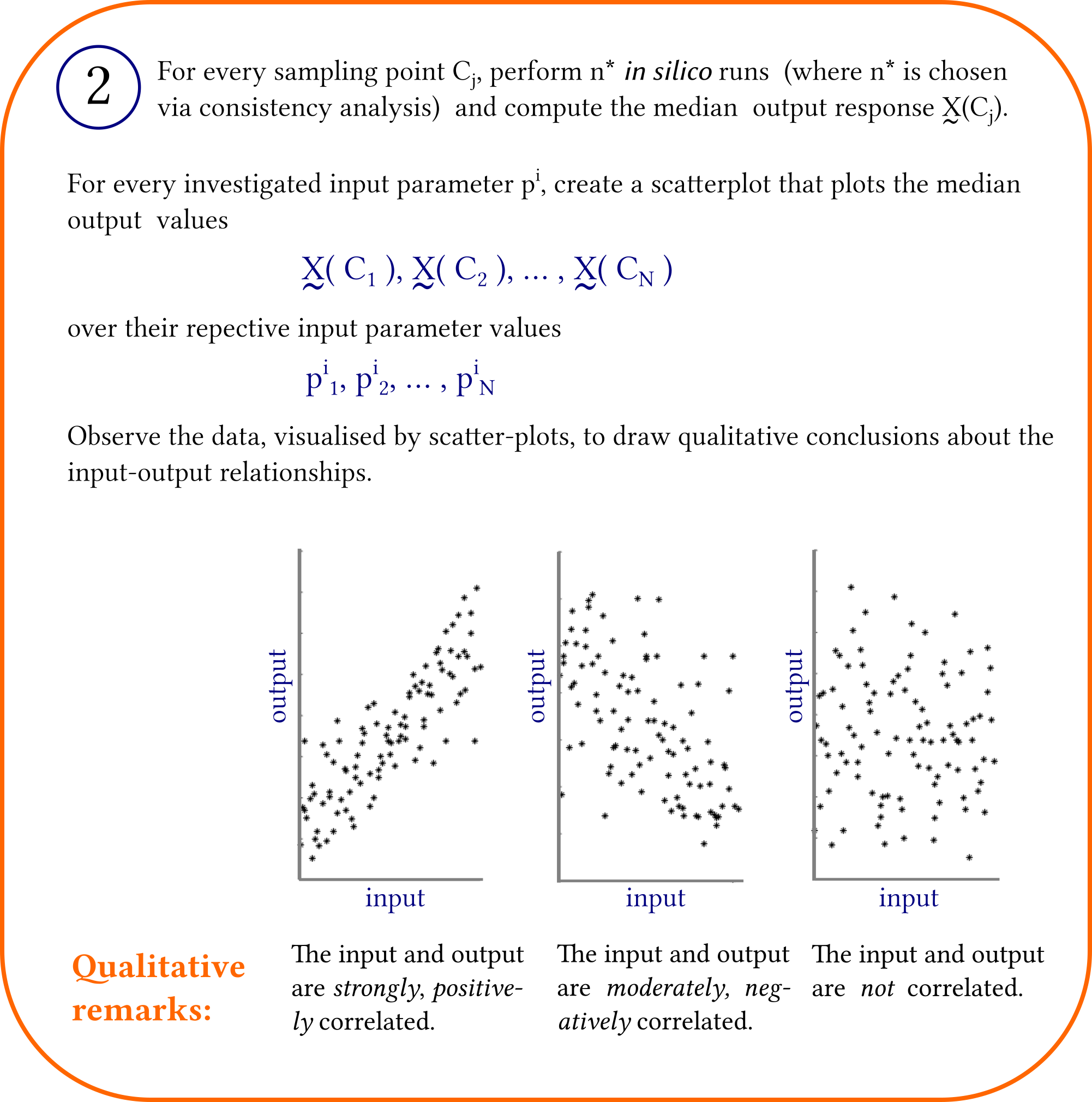}
\\

\includegraphics[scale=0.4]{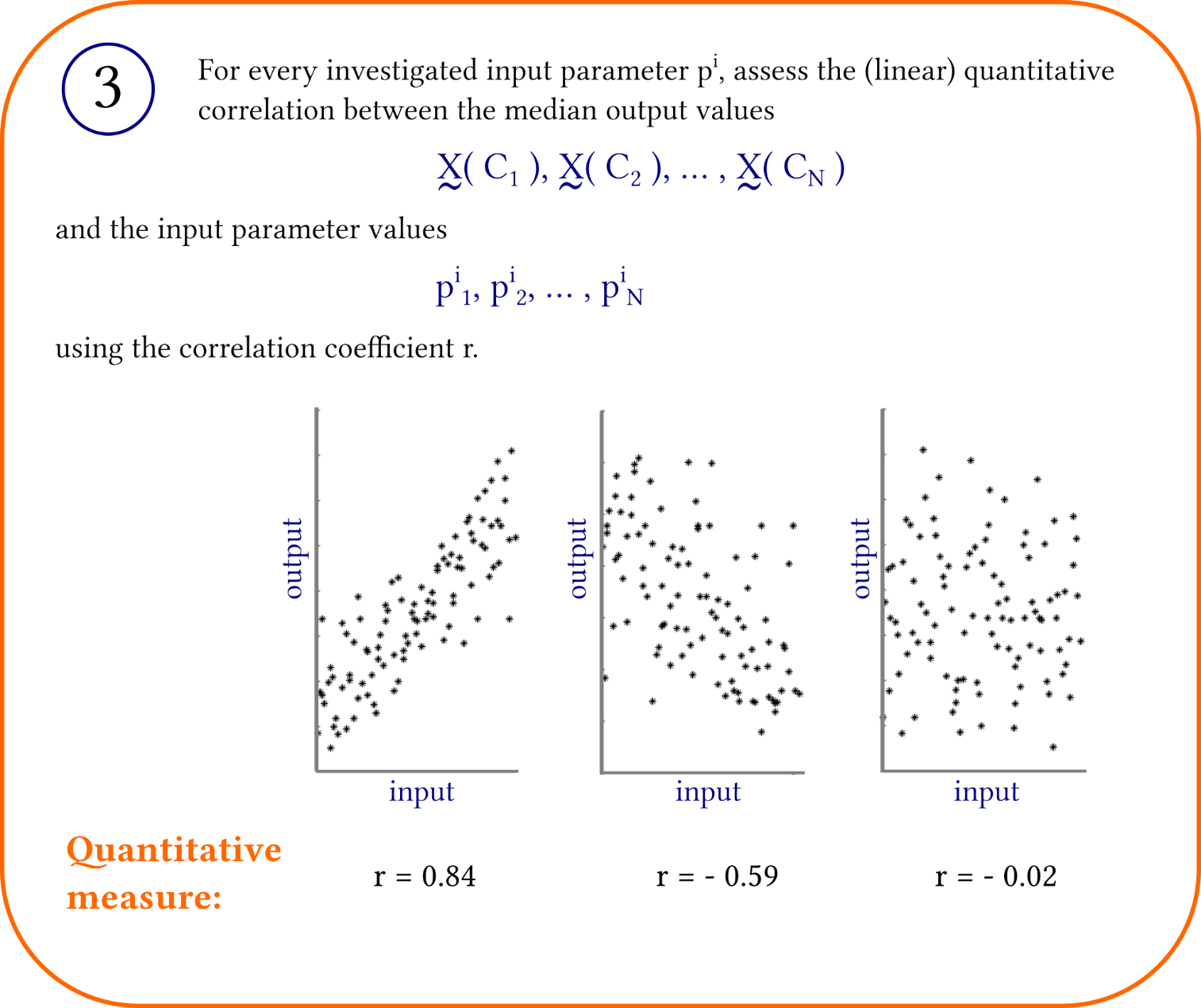}
\\

\includegraphics[scale=0.4]{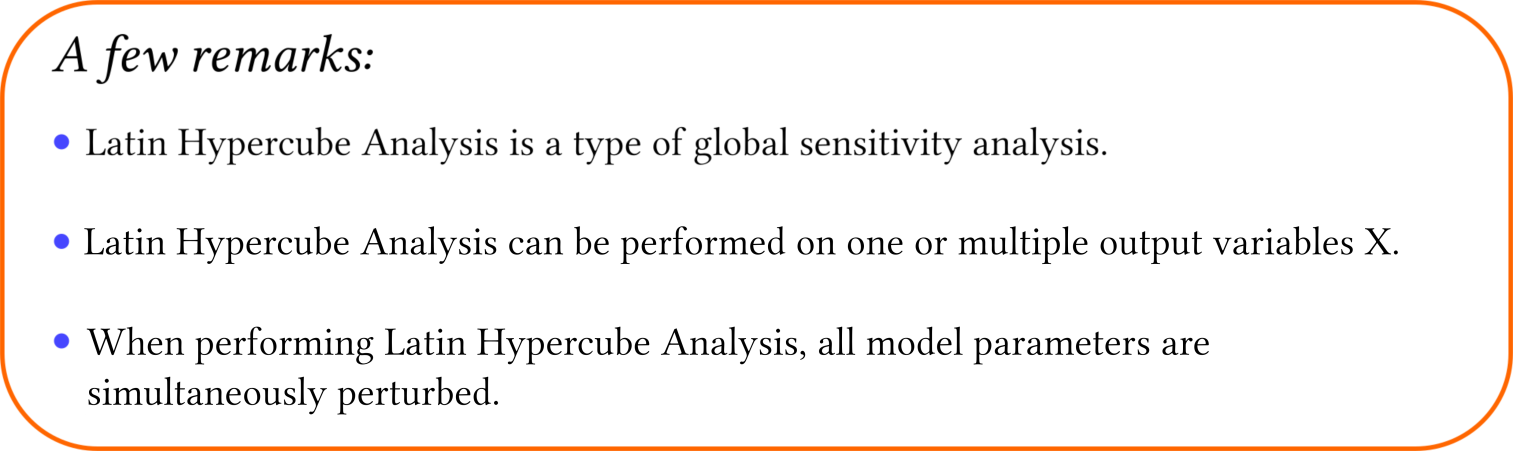}

\newpage
\section{A worked example: Analysing a mathematical cancer model}
\label{sec:workedexample}
\begin{figure}[H]
    \centering
    \includegraphics[scale=0.75]{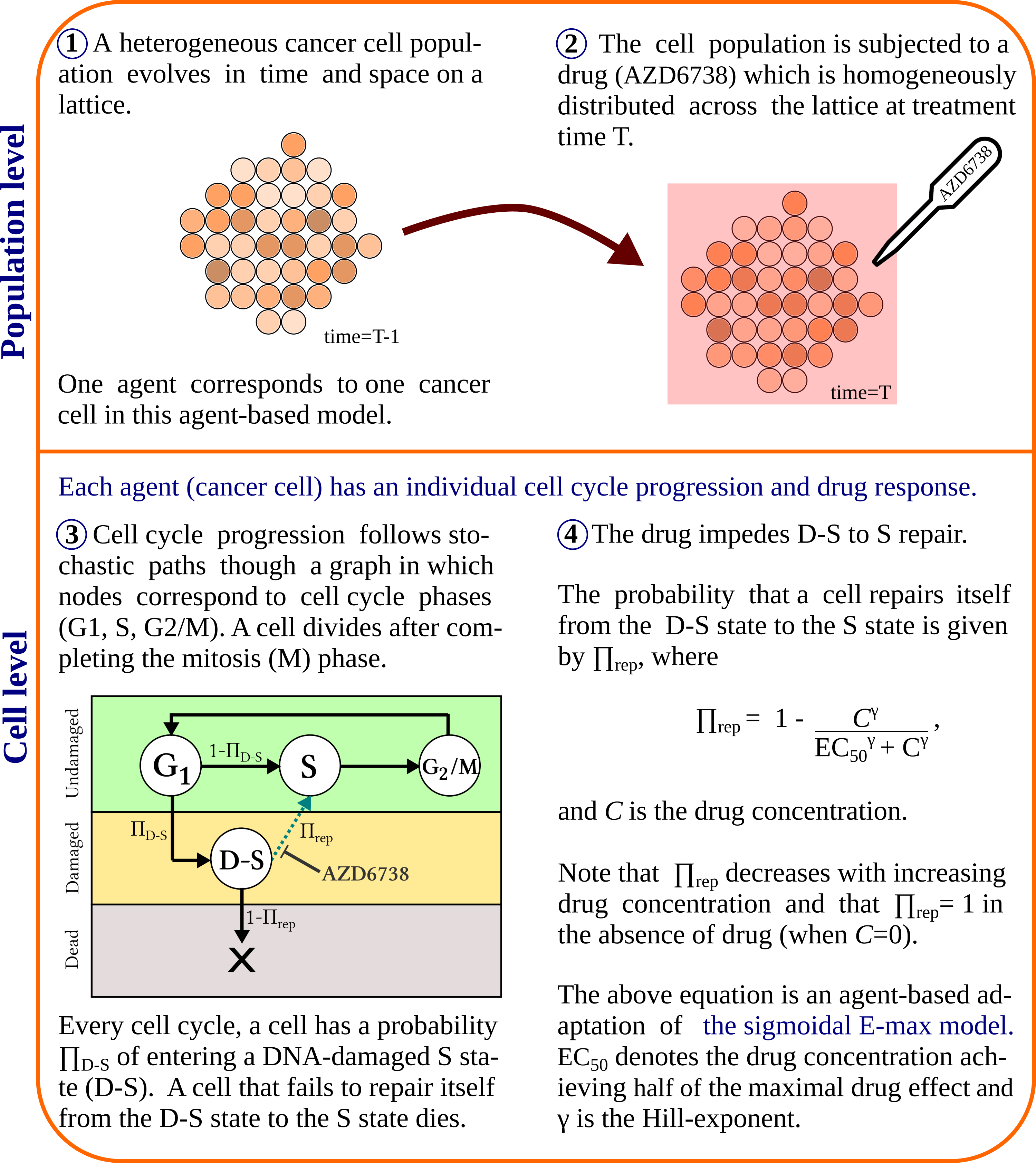}
    \caption{A summary of an agent-based model that is used to simulate human colon carcinoma cells {\it in vitro} subjected to a drug that targets cellular DNA damage responses. 
    In this review, we perform uncertainty and sensitivity analyses on this model as a worked example. Full details are available in the research paper in which this model was first introduced \cite{DDR}.  }
    \label{fig:DDR_model_summary}
\end{figure}

In this section we will perform Consistency Analysis, Robustness Analysis and Latin Hypercube Analysis on an agent-based mathematical model that describes a population of cancer cells ({\it in vitro)} that are subjected to an anti-cancer drug (AZD6738) that may inhibit DNA damage repair in cells and, by extension, cause cell death. 
Full details of this model are available in one of our recent research papers \cite{DDR}, but a pictorial model summary is provided in Figure \ref{fig:DDR_model_summary}. This summary contains sufficient information for our current purposes: performing uncertainty and sensitivity analysis through a worked example. 
In order to do this, we need to specify a set of model inputs and outputs. The full model includes seven input parameters $p^i$, $i=1,..,7$, three of which are mentioned in the model summary and will be investigated in this review. These input parameters are $p^1$: the probability $\Pi_{D-S}$ that a cell enters the damaged S phase in the cell cycle, $p^2$: the drug's EC$_{50}$ value and $p^3$: the Hill-exponent ($\gamma$) used to compute cellular drug responses. Furthermore, we consider two {\it in silico} measurements as outputs, specifically $X^1$: the percentage of DNA-damaged ({\em i.e.} $\gamma$-H2AX positive) cells at the end of the simulation and $X^2$: the cell count at the end of the simulation. 

\newpage 

\subsection{Worked example: Consistency Analysis}
In order to perform Consistency Analysis, we follow steps 1, 2 and 3 outlined in the quick guide in Section \ref{sec:sa_ca_lazy}. 
\\

\textbf{Step 1:} Using the calibrated model parameters, we run our {\it in silico} experiment $20\times(1+5+50+100+300)=9120$ times in order to produce 9120 data samples. Note that, in this case, we have two output responses of interest, and thus one data sample consists of an output-pair ($X^1, X^2$). 
Post the {\it in silico} production of data, we organise our data samples into five groups of distributions, where each group consists of 20 distributions of data samples. 
In the first group, each one of the 20 distributions includes only one data sample. In the second, third, forth and fifth group, each distribution respectively includes 5, 50, 100 or 300 data samples. 
\\

\textbf{Step 2:} In each distribution group, we compute and plot the $\hat{A}$-measure (in original and scaled form) for each of its 20 distributions, as is done in Figures \ref{fig:ddr_sa_ca1} through to  \ref{fig:ddr_sa_ca300}. These figures clearly demonstrate that the statistical significance decreases with increasing distribution size $n$. 
Note that the two output responses of interest are computed and plotted independently of each other, as we are aiming to find a distribution size that yields a small statistical significance for both $X^1$ and $X^2$. 
\\

\textbf{Step 3:} The largest scaled $\hat{A}$-measure in each distribution group is computed and plotted over the group's distribution size. The smallest distribution size for which the statistical significance is small ({\em i.e}. $\leq$ 0.56) for both $X^1$ and $X^2$ is denoted $n^*$. 
By regarding Table \ref{tab:ddr_sa_ca_maxAvals} and Figure \ref{fig:ddr_sa_ca_tot}, we can see that, in this case, $n^*=100$.  Accordingly, we determine that $100$ simulation runs are sufficient to mitigate uncertainty originating from intrinsic model stochasticity. 
Thus when talking about, for example, average values and standard deviations produced by this model, we should base these measures on data samples from $100$ {\it in silico} runs.  

\begin{table}[H]
\begin{center}
\begin{tabular}{|x{2.5cm}||x{1.2cm}|x{1.2cm}|x{1.2cm}|x{1.2cm}|x{1.2cm}|x{1.2cm}|x{1.2cm}|}
\hline
       \diaghead(-3,2){\hskip \hsize}{output}{distribution \\ size}& n=1 & n=5 & n=50 & n=100 & n=300 \\ \hline
    $X^1$& 1 & 0.92 & 0.61 & 0.55 & 0.54\\
    $X^2$ & 1 &	 0.84 & 0.59 & 0.56 & 0.54 \\
\hline
\end{tabular}
\end{center}
\caption{Maximal scaled $\hat{A}$-values for various distribution sizes $n$. The output responses are $X^1$: the percentage of $\gamma$H2AX-positive ({\em i.e.} DNA-damaged) cells, and $X^2$: the cell count at the end of the simulation.}
\label{tab:ddr_sa_ca_maxAvals}
\end{table}
\normalsize

\begin{center}

\begin{figure}[H]
    \centering
    \includegraphics[scale=0.65]{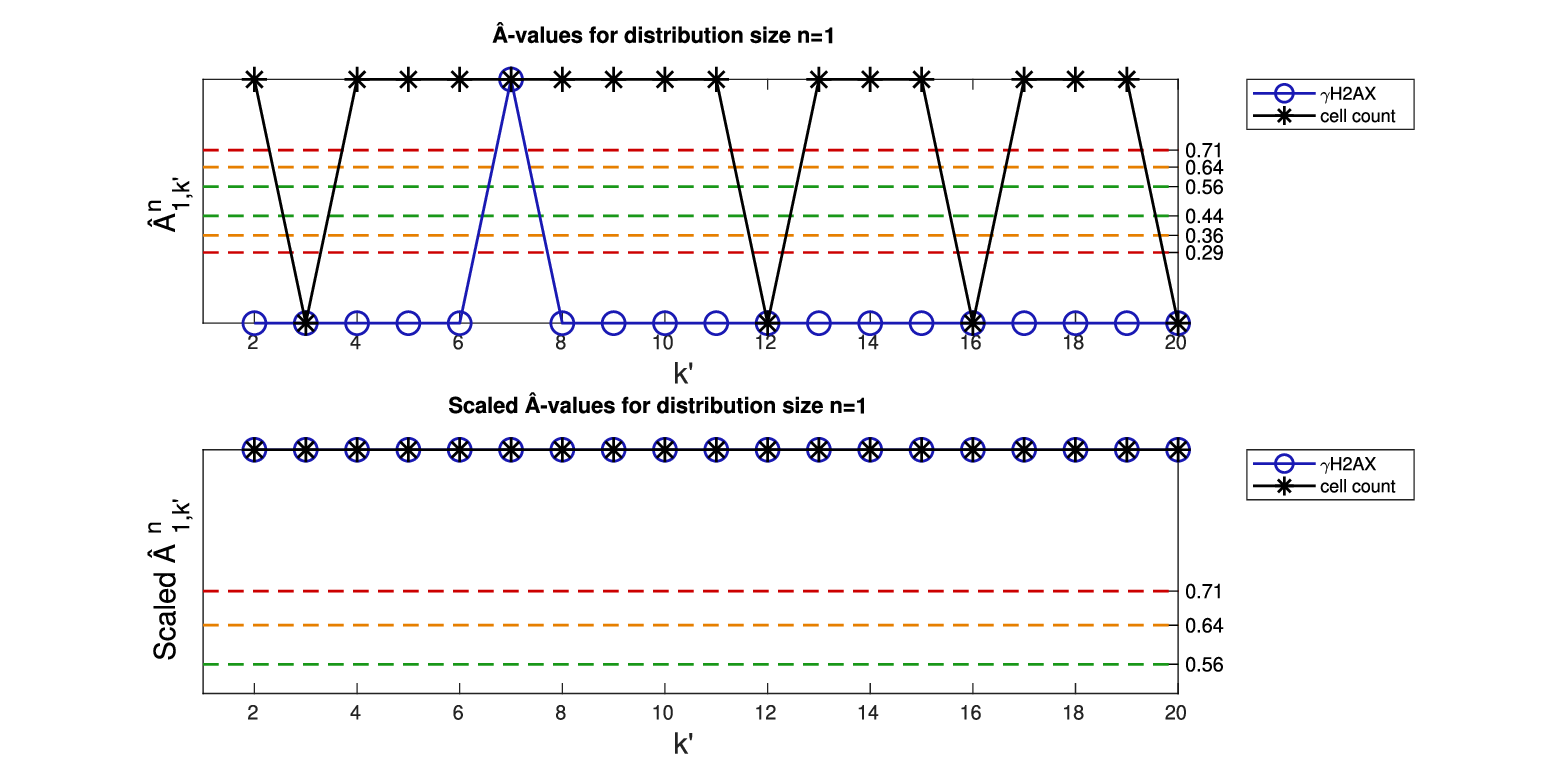}
    \caption{Consistency Analysis,  $\hat{A}$-values in initial (top) and scaled (bottom) form for distribution size $n=1$. }
    \label{fig:ddr_sa_ca1}
\end{figure}

\begin{figure}[H]
    \centering
    \includegraphics[scale=0.65]{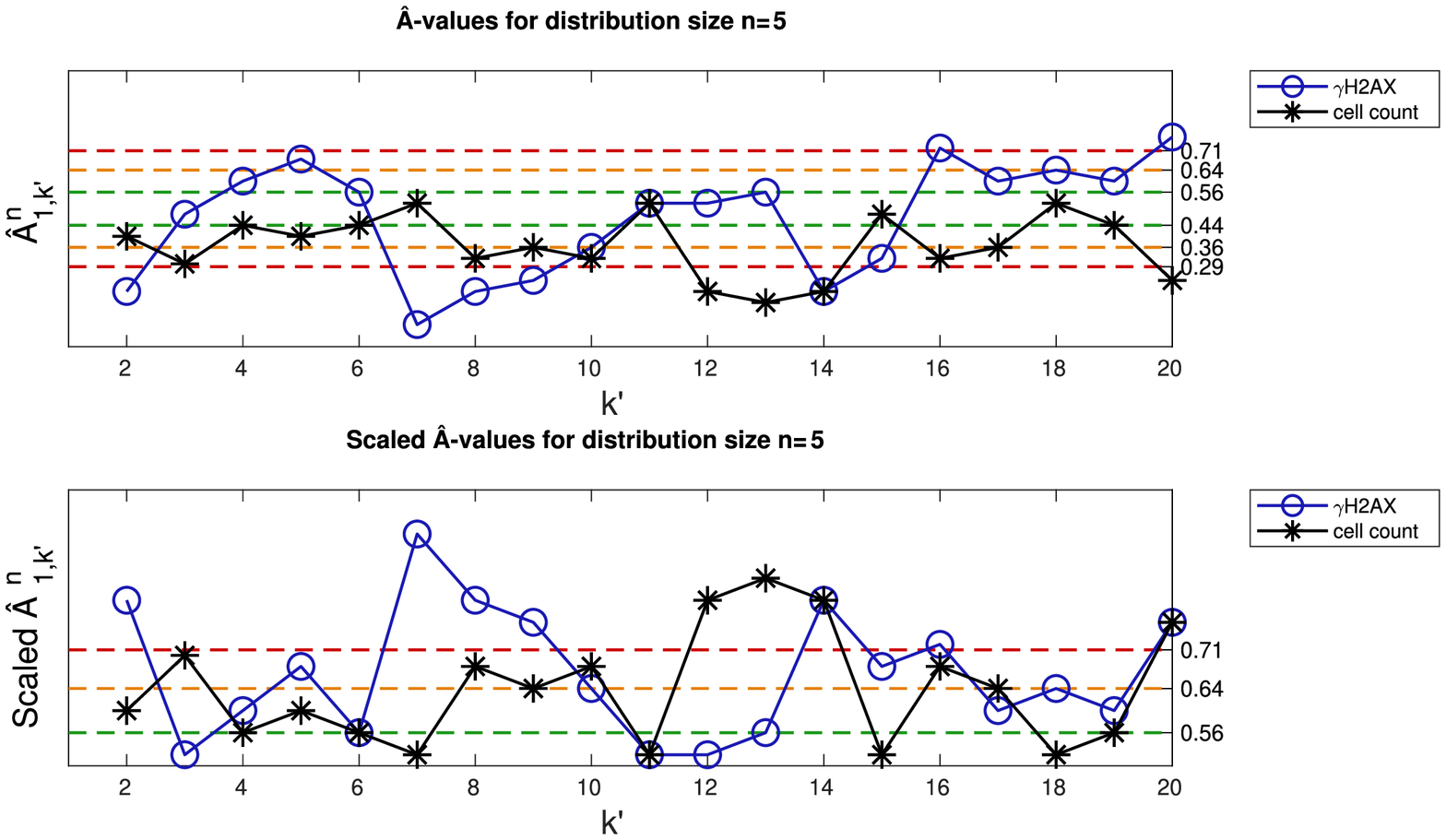}
    \caption{Consistency Analysis, $\hat{A}$-values in initial (top) and scaled (bottom) form  for distribution size $n=5$. }
    \label{fig:ddr_sa_ca5}
\end{figure}

\begin{figure}[H]
    \centering
    \includegraphics[scale=0.65]{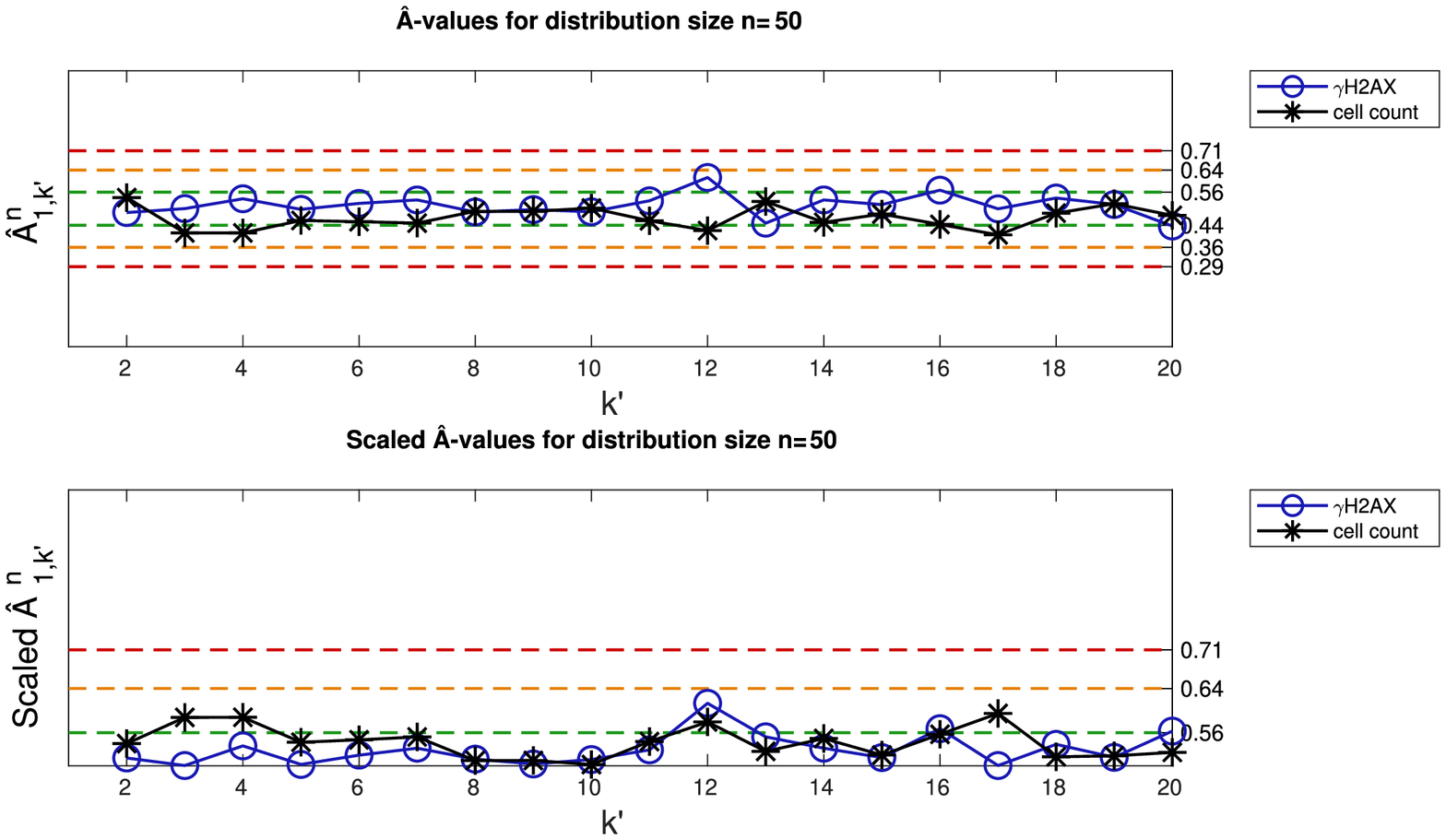}
    \caption{Consistency Analysis, $\hat{A}$-values in initial (top) and scaled (bottom) form  for distribution size $n=50$. }
    \label{fig:ddr_sa_ca50}
\end{figure}

\begin{figure}[H]
    \centering
    \includegraphics[scale=0.65]{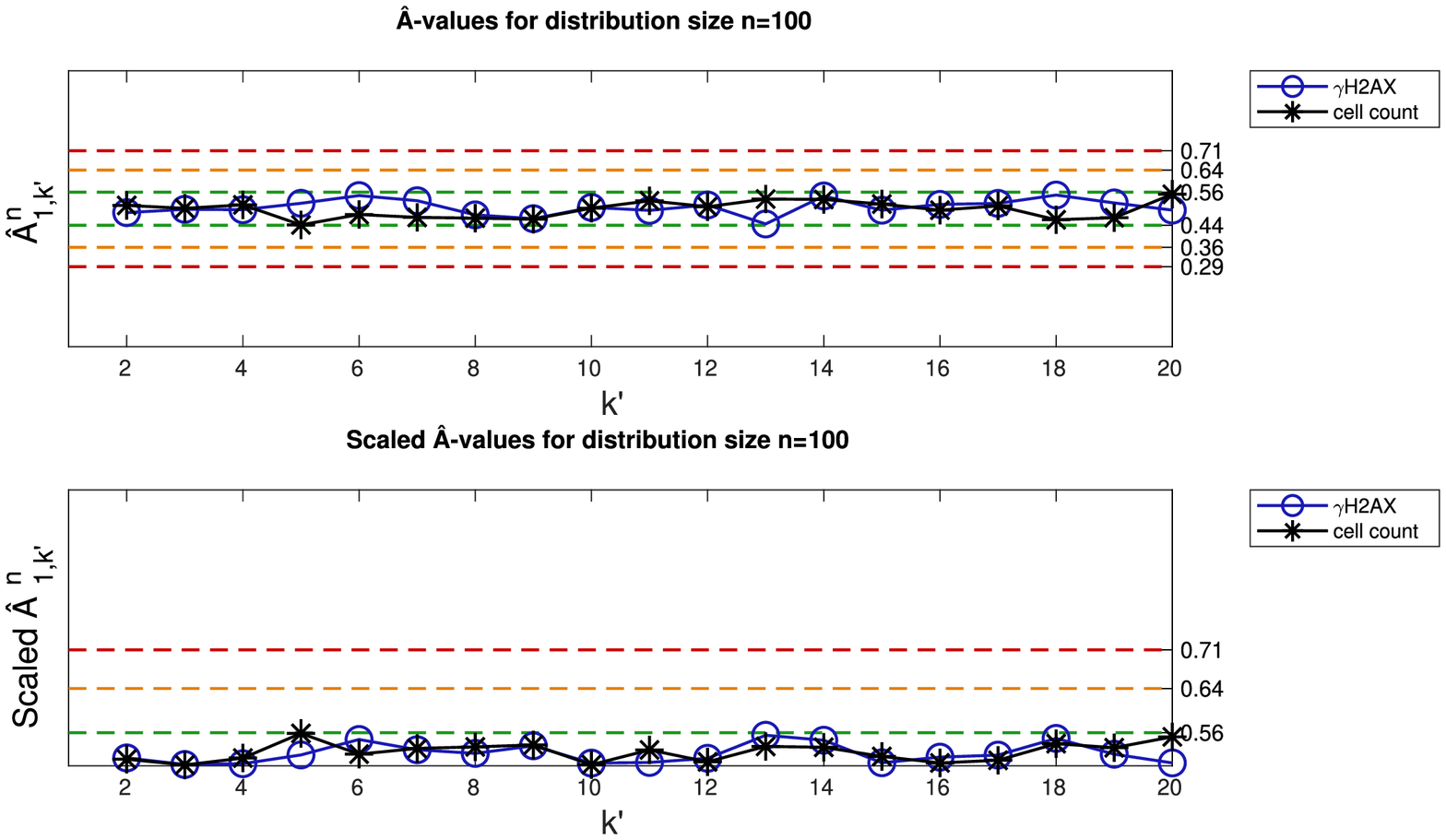}
    \caption{Consistency Analysis, $\hat{A}$-values in initial (top) and scaled (bottom) form  for distribution size $n=100$. }
    \label{fig:ddr_sa_ca100}
\end{figure}

\begin{figure}[H]
    \centering
    \includegraphics[scale=0.65]{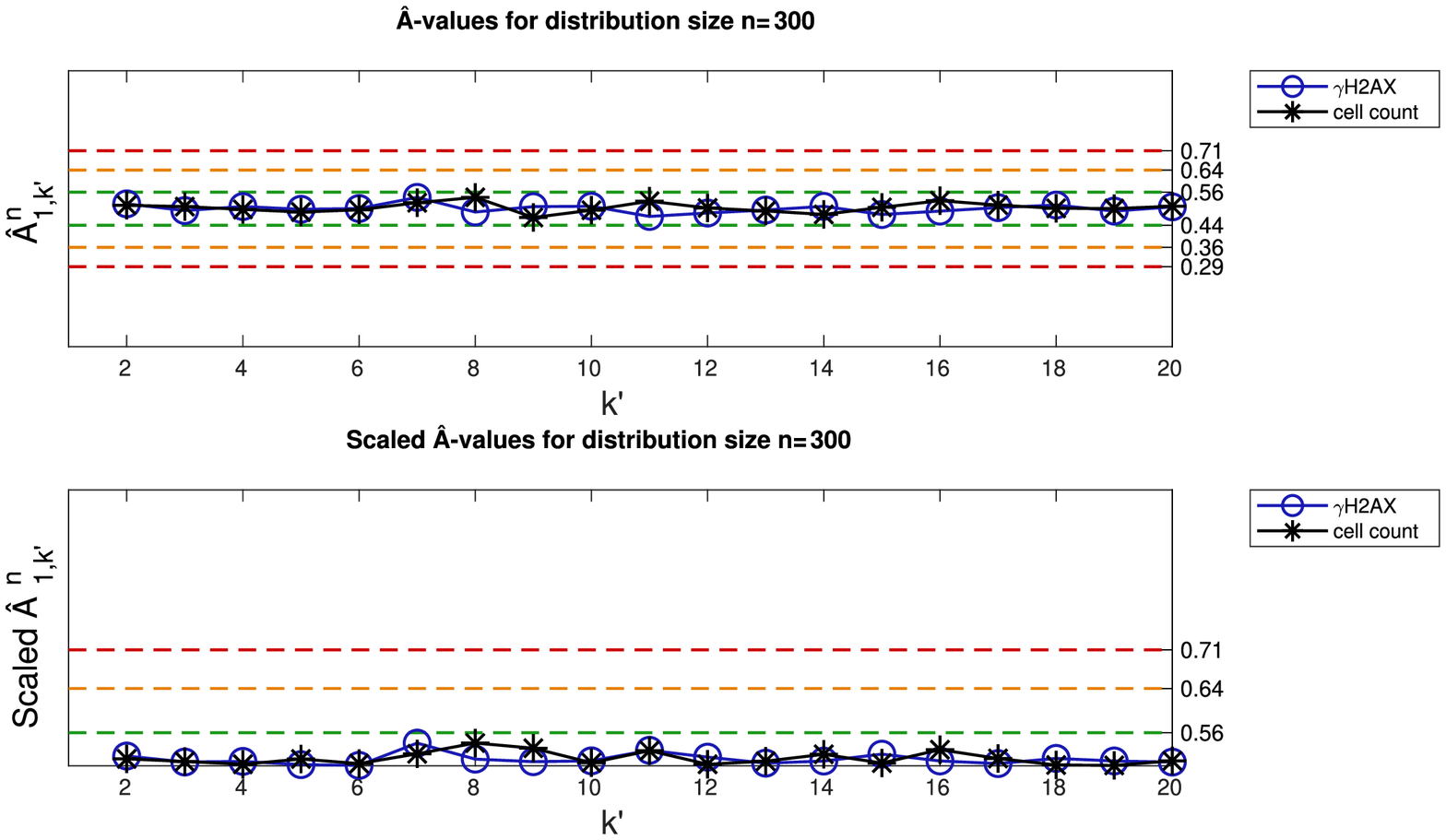}
    \caption{Consistency Analysis, $\hat{A}$-values in initial (top) and scaled (bottom) form  for distribution size $n=300$. }
    \label{fig:ddr_sa_ca300}
\end{figure}

\begin{figure}[H]
    \centering
    \includegraphics[scale=0.53]{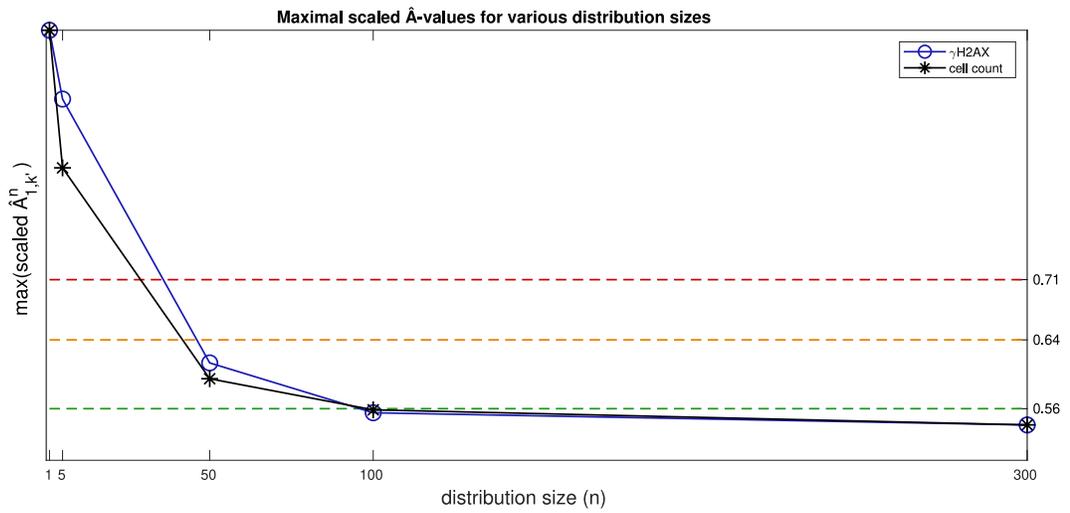}
    \caption{Consistency Analysis, maximal scaled $\hat{A}$-values for various distribution sizes $n$. }
    \label{fig:ddr_sa_ca_tot}
\end{figure}
\end{center}

\newpage 

\subsection{Worked example: Robustness Analysis}
We now set out to perform Robustness Analysis using steps 1, 2 and 3 described in the quick guide in Section \ref{sec:sa_qg_ra}.  
Here, our goal is to  investigate how sensitive model output responses are to local (one at a time) parameter perturbations of the inputs $\Pi_{D-S}$, $EC_{50}$ and $\gamma$. 
\\

\textbf{Step 1:} The first thing we must do here is to decide appropriate parameter ranges to investigate for each input parameter. This decision should ideally be guided by information from experimental data or biological/mathematical knowledge about the modelling scenario at hand. The mathematical model summarised in Figure \ref{fig:DDR_model_summary} is driven by {\it in vitro} data, and from this data suitable parameter ranges can be deduced and established (full details are available in the original model paper \cite{DDR}). 

For example, recall that the parameter $EC_{50}$ corresponds to the drug concentration $C$ that achieves half of $E_{max}$, the maximal drug effect. 
Now the {\it in vitro} data shows that $C=1\mu$M yields $\approx 0.5E_{max}$ and that $C=3\mu$M (or higher) yields $\approx E_{max}$, whilst a drug concentration of $C=0.3\mu$M evokes roughly the same drug response as the control case (when no drug at all is applied and thus $C=0$) \cite{DDR, Checkley2015}. 
From this, we can reason that it is appropriate to investigate $EC_{50}$ parameter values in the range $(1 \pm 0.75)\mu$M. Similarly, we may decide to investigate $\Pi_{D-S}$ values in the parameter range $(75 \pm 10)$\%, and $\gamma$ values in $(2 \pm 1)$.

Now that we have established our parameter ranges of interest, it remains to decide how many (here evenly spaced) parameter values in these ranges to test. 
The more parameter values we test, the more detailed information we get, but recall that for every investigated parameter we need to create $n^*$ data samples (where $n^*$ is determined in the Consistency Analysis). 
The decision of how many parameter values to include in the Robustness Analysis should be informed by the fineness of available {\it in vitro/in vivo} data, computational costs, scientific reasoning and model application. 
If a model is being used in a pharmaceutical setting, for example, a detailed analysis may be of extra importance. Here, we respectively choose to investigate 9,7 and 9 evenly spaced parameter values (including the calibrated parameter values) in the parameter ranges for $\Pi_{D-S}$, $EC_{50}$ and $\gamma$. Thus we need to produce $n^* \times (9+7+9) = 100 \times (25)$ {\it in silico} data samples. 
\\

\begin{figure}[H]
    \centering
    \includegraphics[scale=0.5]{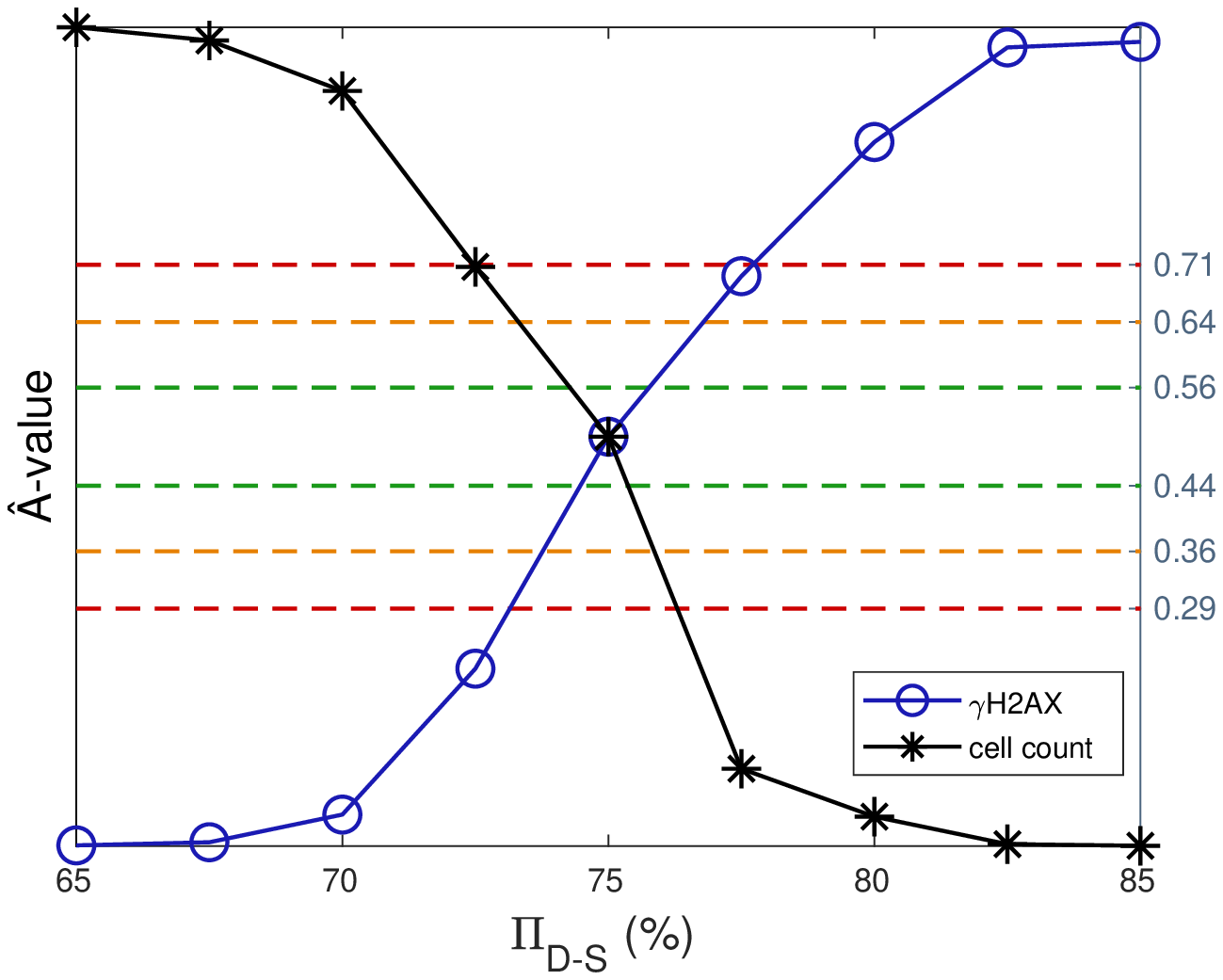}
    \includegraphics[scale=0.5]{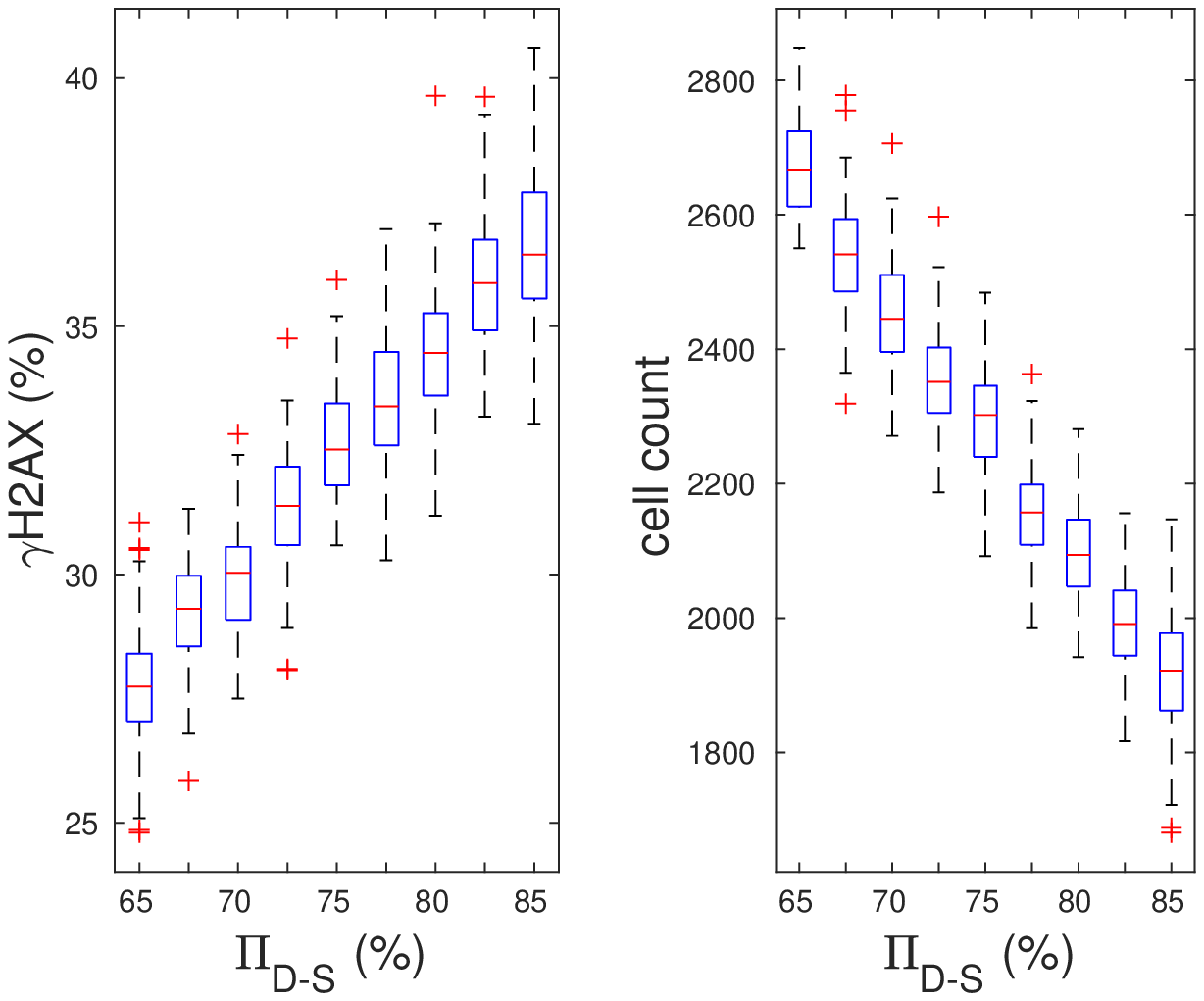}
    \caption{Robustness Analysis, 
    Left: $\hat{A}$-values resulting from comparisons between distributions of data samples produced with perturbed $\Pi_{D-S}$ values, and the distribution produced with the calibrated (unperturbed) $\Pi_{D-S}$ value. 
    Right: Output responses, in terms of percentage of $\gamma$H2AX-positive ({\em i.e.} damaged) cells, and cell count as a result of perturbations to the input variable $\Pi_{D-S}$.}
    \label{fig:ddr_sa_ra3}
\end{figure}

\begin{figure}[H]
    \centering
    \includegraphics[scale=0.5]{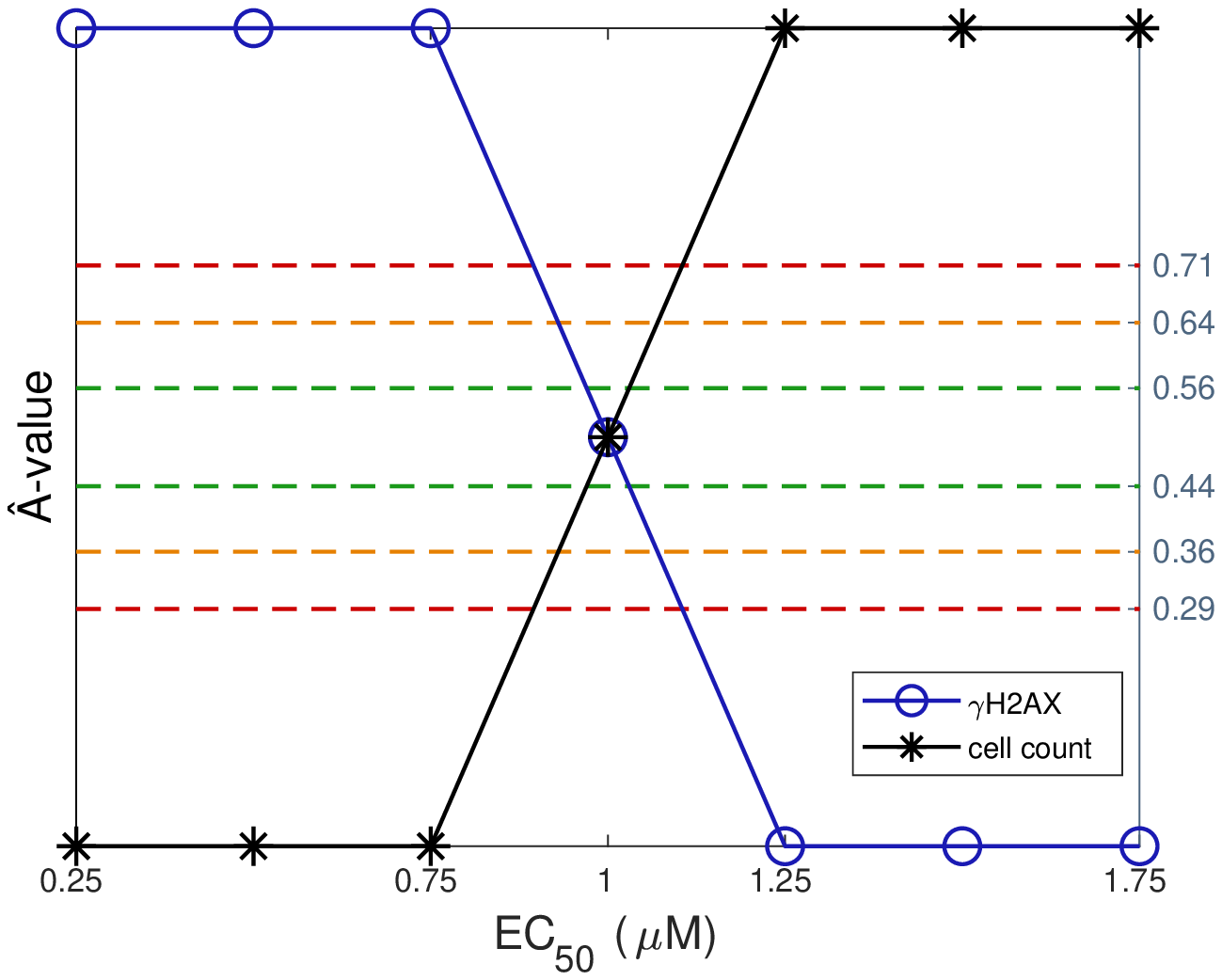}
    \includegraphics[scale=0.5]{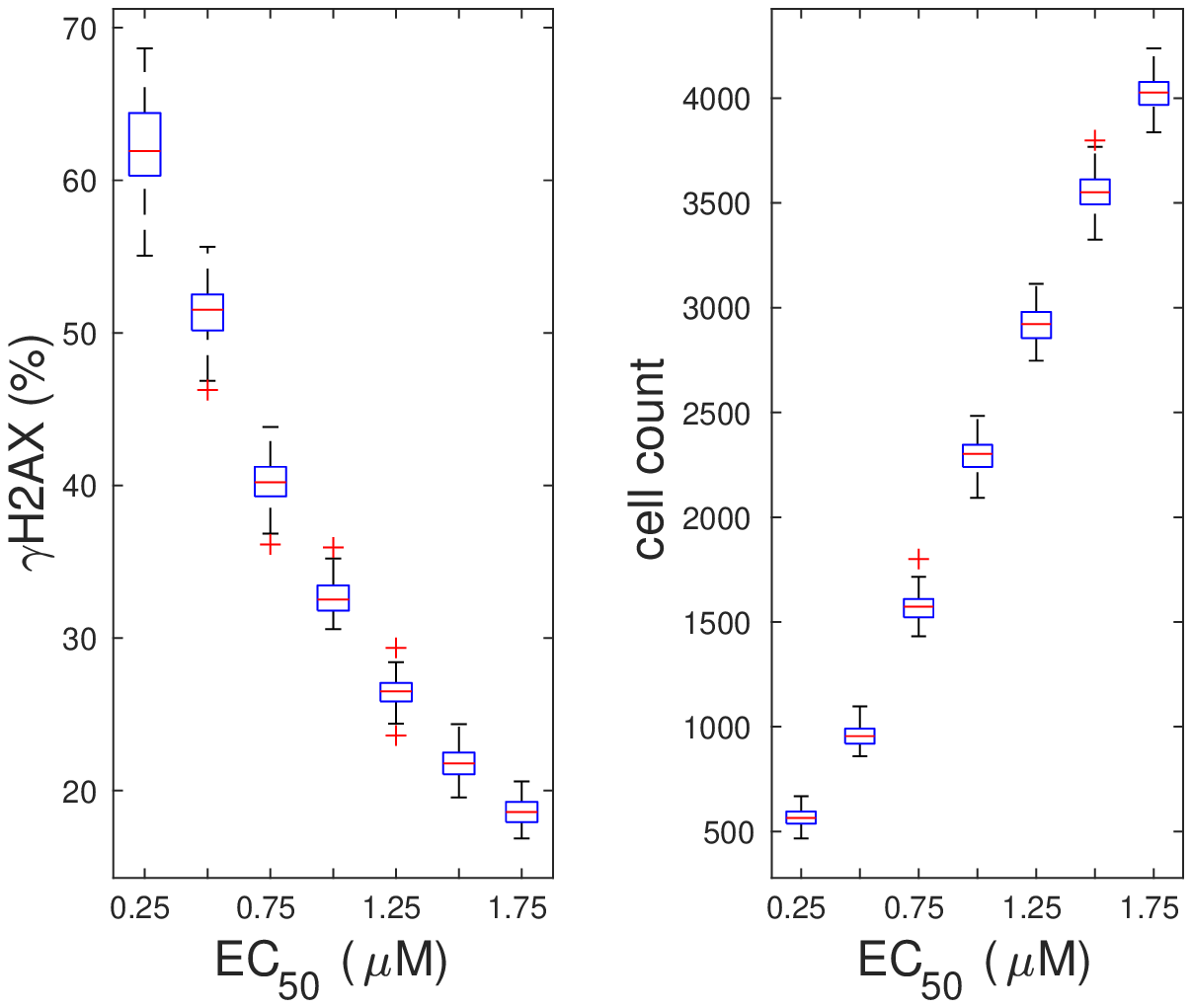}
    \caption{Robustness Analysis, Left:   $\hat{A}$-values resulting from comparisons between distributions of data samples produced with perturbed $EC_{50}$ values, and the distribution produced with the calibrated (unperturbed) $EC_{50}$ value.
    Right: Output responses, in terms of percentage of $\gamma$H2AX-positive ({\em i.e.} damaged) cells, and cell count as a result of perturbations to the input variable $EC_{50}$. 
   }
    \label{fig:ddr_sa_ra5}
\end{figure}

\begin{figure}[H]
    \centering
    \includegraphics[scale=0.5]{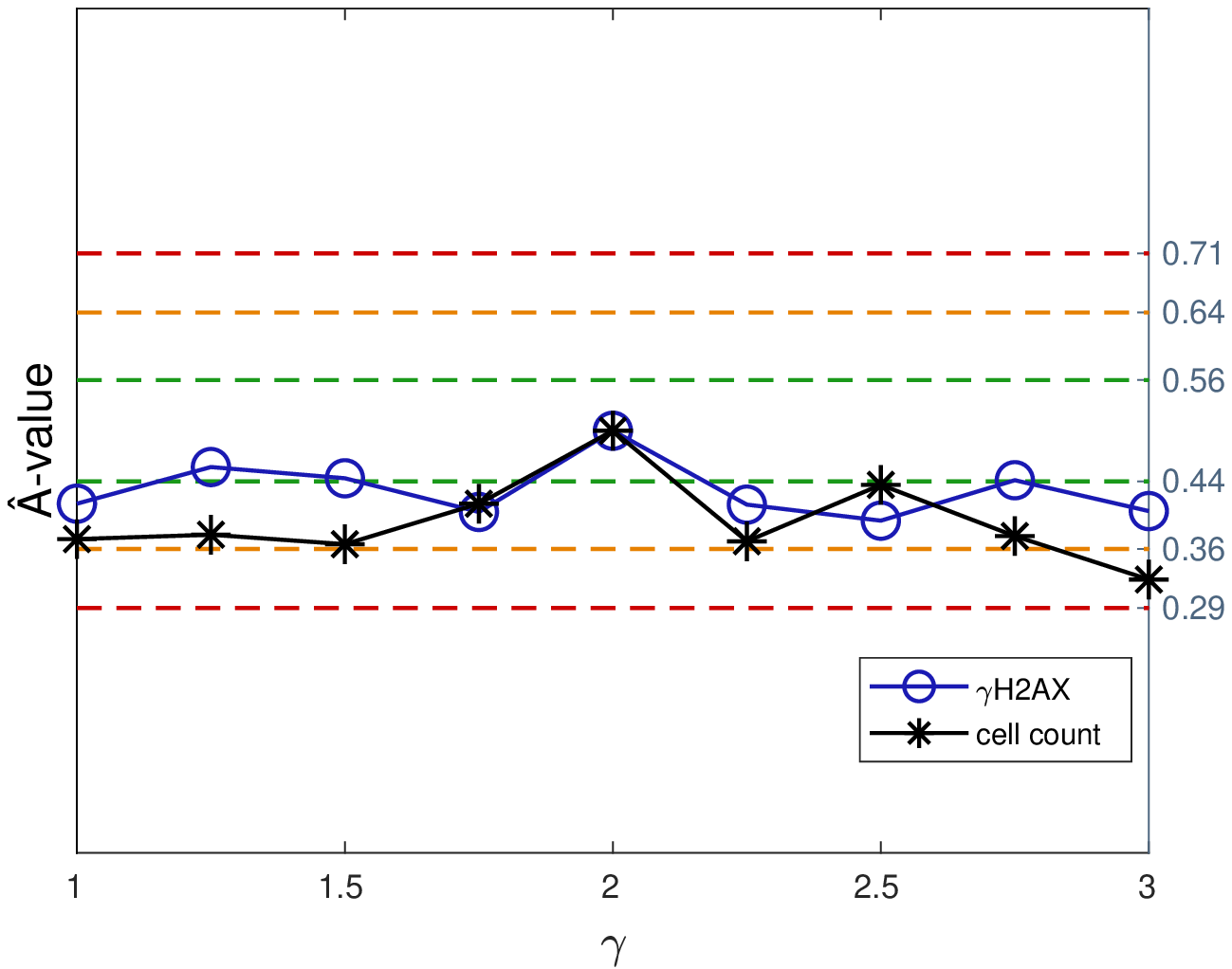}
    \includegraphics[scale=0.5]{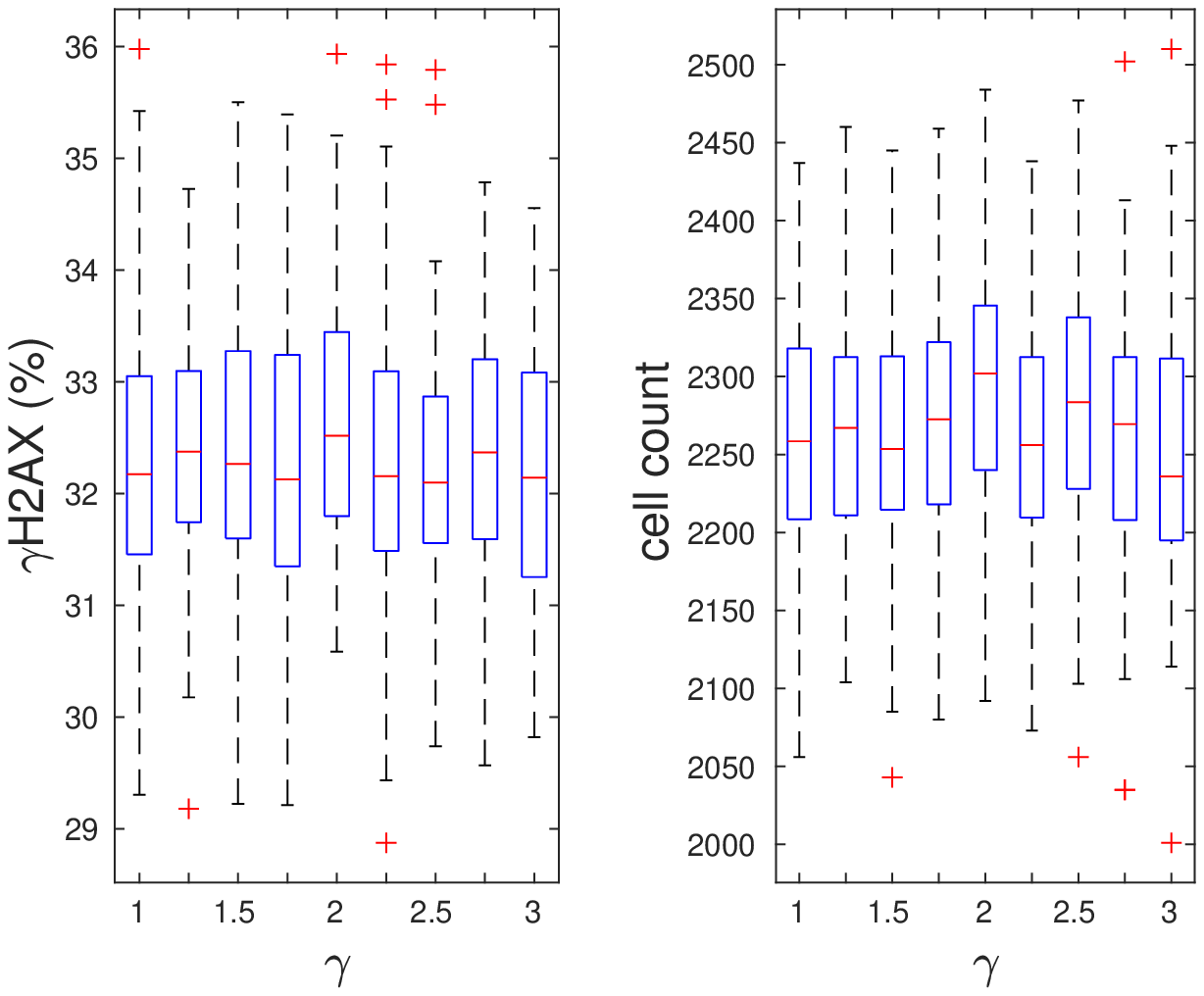}
    \caption{Robustness Analysis, Left:  $\hat{A}$-values resulting from comparisons between distributions of data samples produced with perturbed $\gamma$ values, and the distribution produced with the calibrated (unperturbed) $\gamma$ value.
    Right: Output responses, in terms of percentage of $\gamma$H2AX-positive ({\em i.e.} damaged) cells, and cell count as a result of perturbations to the input variable $\gamma$.
    }
    \label{fig:ddr_sa_ra6}
\end{figure}

\textbf{Steps 2 and 3:}  Post {\it in silico} simulations, the $\hat{A}$-measures are computed and plotted over all parameter values $p^i_j$ for each input parameter $p^i$. Such plots are here shown for input parameters $\Pi_{D-S}$ (Figure \ref{fig:ddr_sa_ra3}), $EC_{50}$ (Figure \ref{fig:ddr_sa_ra5}) and $\gamma$ (Figure \ref{fig:ddr_sa_ra6}).     

The data samples produced in the {\it in silico} experiments are also represented using boxplots in these figures. The $\hat{A}$-measures and the boxplots demonstrate the effect that local perturbations of the input parameters have on both output responses $X^1$ and $X^2$. 

Figure \ref{fig:ddr_sa_ra3} illustrates that increasing the probability $\Pi_{D-S}$ that a cell enters the damaged S state increases the percentage of $\gamma$H2AX-positive cells and decreases the cell count, as is to be expected. Further, Figure \ref{fig:ddr_sa_ra5} demonstrates that the model output is highly sensitive to perturbations of $EC_{50}$. Increasing $EC_{50}$ results in a higher percentage of $\gamma$H2AX-positive cells and a lower cell count. Finally, Figure \ref{fig:ddr_sa_ra6} shows that the regarded output responses are less sensitive to small perturbations of the Hill-exponent $\gamma$ than to small perturbations of $\Pi_{D-S}$ and $EC_{50}$.

\newpage 

\subsection{Worked example: Latin Hypercube Analysis}
Following steps 1, 2 and 3, as described in the quick guide in Section \ref{sec:sa_lhc_lazy}, Latin Hypercube Analysis is here performed in order to investigate how sensitive output responses are to {\it global} parameter perturbations. 
We here investigate parameter values within parameter ranges that we consider to be `plausible' post Robustness Analysis. 
\\

\textbf{Step 1:} 
For each input parameter, we decide to split the investigated parameter range into $N=100$ intervals. Why 100 you might ask? Well, in the original paper where we first introduced this model \cite{DDR}, the model took seven input parameters and thus $q=7$. Accordingly, we tried using $N=4q/3\approx10$ and $N=2q=14$ intervals at first (following the suggestions discussed in Section \ref{sec:sa_lhc}) but neither of these options produced enough data samples to yield meaningful information in steps 2 and 3 below. Therefore, we decided to use $N=100$ instead, as this choice covered a larger range of the input parameter space whilst coming at a feasible computational cost. Now, we can use the built-in \textsf{MATLAB} function \textsf{lhsdesign} \cite{MATLAB:2019b} to create combinations of input parameter values (represented by a point $C_p$ in input parameter space) that shall be used to produce the {\it in silico} data samples needed for Latin Hypercube Analysis. 

\begin{figure}[H]
    \centering
    \includegraphics[scale=0.55]{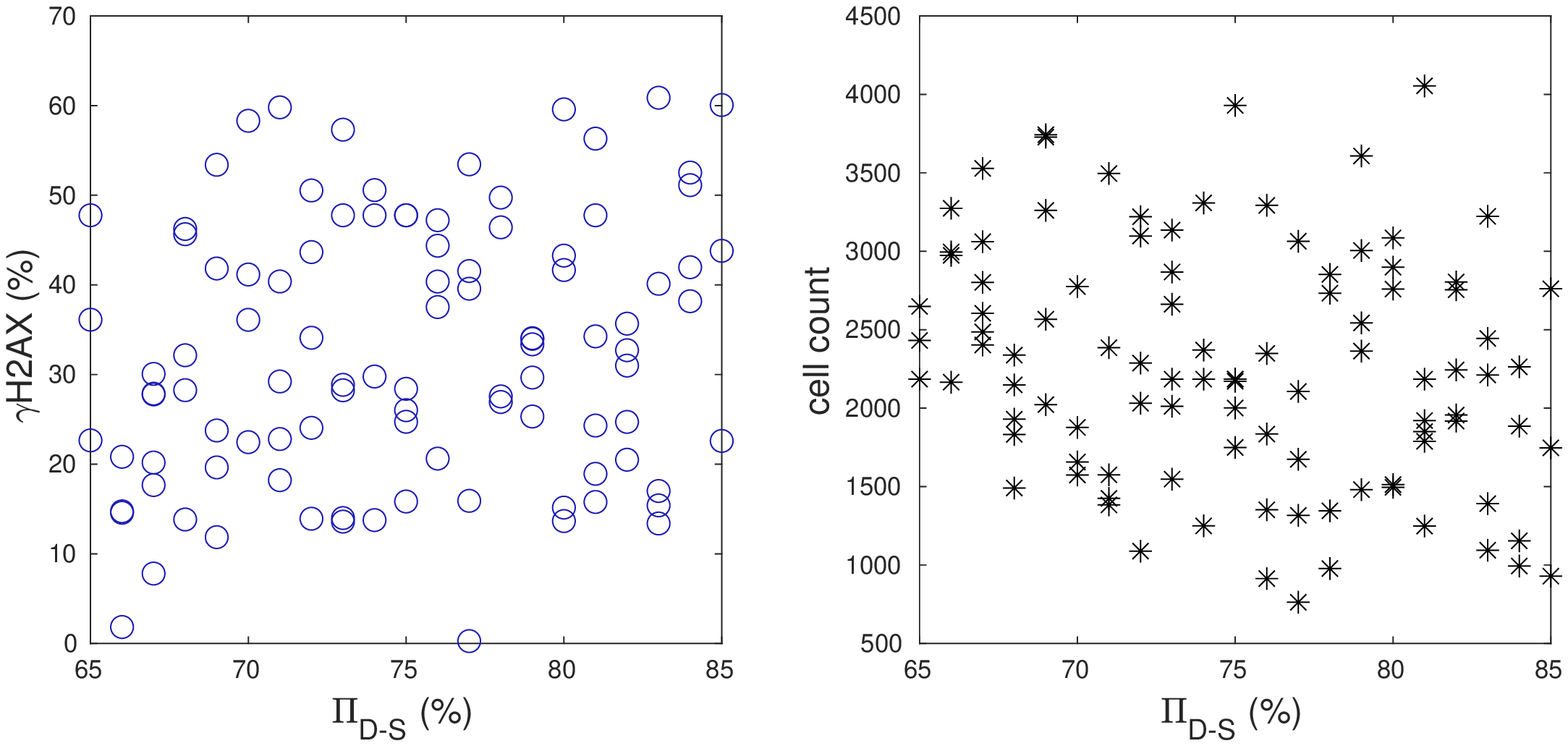}
    \caption{Latin Hypercube Analysis. Output responses in terms of $\gamma$H2AX-positive cells (left) and cell count (right) when global parameter perturbations are performed. The scatterplots show the correlation between outputs and the input value of $\Pi_{D-S}$. }
    \label{fig:ddr_sa_lhc3}
\end{figure}

\begin{figure}[H]
    \centering
    \includegraphics[scale=0.55]{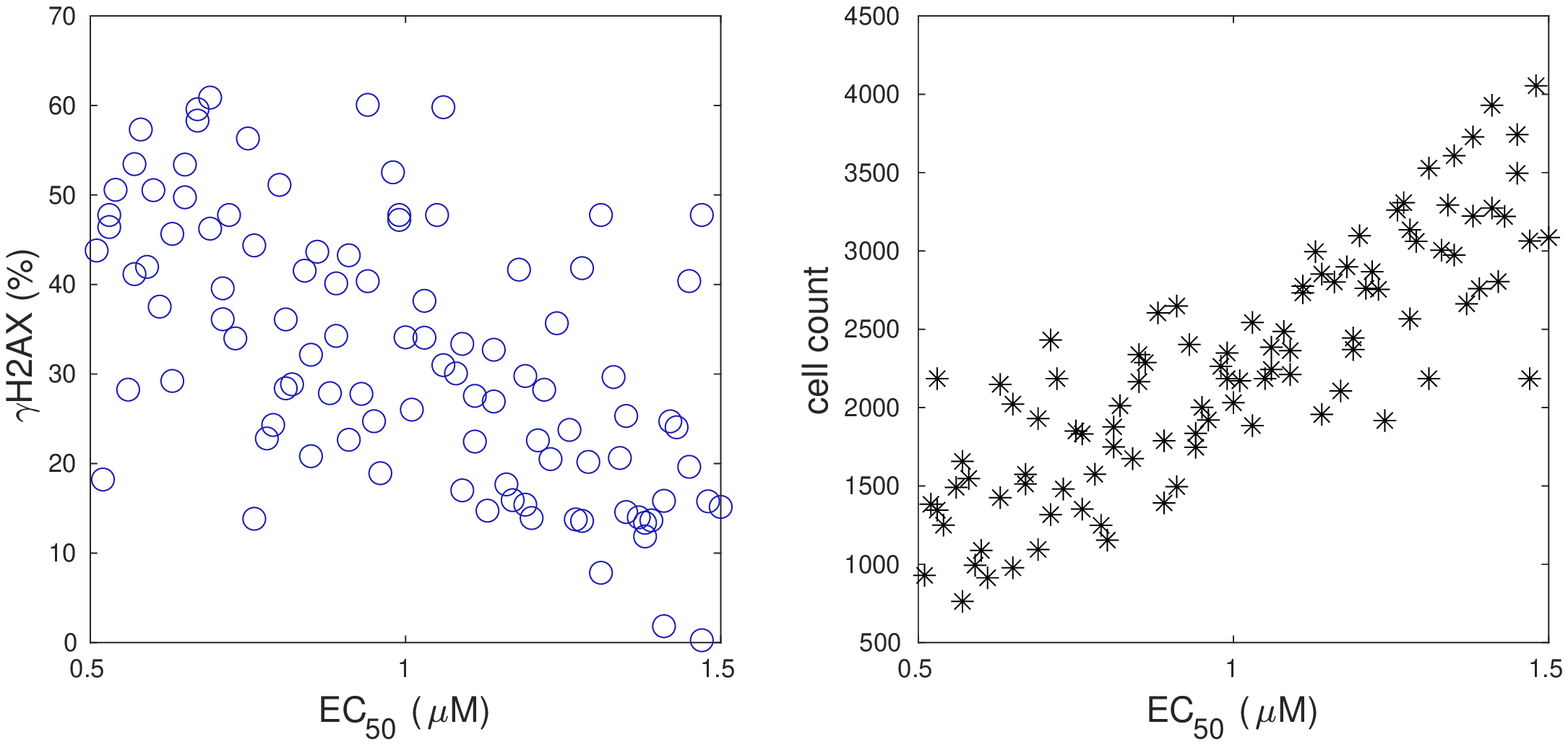}
    \caption{Latin Hypercube Analysis. Output responses in terms of $\gamma$H2AX-positive cells (left) and cell count (right) when global parameter perturbations are performed. The scatterplots show the correlation between outputs and the input value of $EC_{50}$. }
    \label{fig:ddr_sa_lhc5}
\end{figure}

\begin{figure}[H]
    \centering
    \includegraphics[scale=0.55]{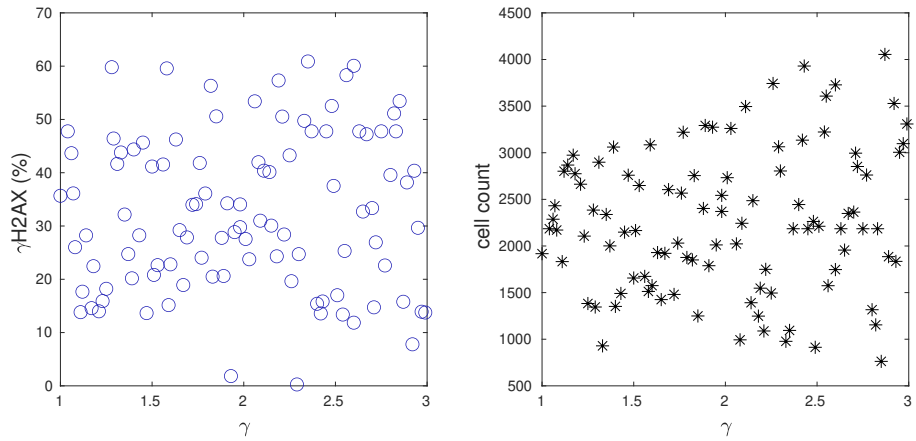}
    \caption{Latin Hypercube Analysis. Output responses in terms of $\gamma$H2AX-positive cells (left) and cell count (right) when global parameter perturbations are performed. The scatterplots show the correlation between outputs and the input value of $\gamma$.}   \label{fig:ddr_sa_lhc6}
\end{figure}

\textbf{Step 2:}  
For each $C_p$, median output responses ($X^1$ and $X^2$) of $n^*=100$ {\it in silico} runs are computed. These are plotted in {\it 'output-over-input'} scatterplots for each investigated input parameter in Figures \ref{fig:ddr_sa_lhc3}, \ref{fig:ddr_sa_lhc5} and \ref{fig:ddr_sa_lhc6}. From these figures we can make some qualitative remarks: Figure \ref{fig:ddr_sa_lhc3} indicates that the relationships between the input variable $\Pi_{D-S}$ and the output responses $X^1$ and $X^2$ are, respectively, positively and negatively correlated. This agrees with the intuitive notion that if the probability that a cell enters the D-S state increases, so does the percentage of damaged cells ($X^1$) whilst the cell count ($X^2$) decreases as more cells will be susceptible to the drug and potentially die. 
The scatterplots in Figure \ref{fig:ddr_sa_lhc5} demonstrate that the input variable $EC_{50}$ impacts the output responses more than do other investigated input parameters (within the regarded ranges). $EC_{50}$ is negatively, linearly correlated with $X^1$ and positively, linearly correlated with $X^2$.
In Figure \ref{fig:ddr_sa_lhc6}, however, there is no visually apparent correlation between the input parameter $\gamma$ and the output.

\textbf{Step 3:} 
After making some qualitative remarks in Step 2, we now compute the Pearson Product Moment Correlation Coefficients between the various input-output pairs for a quantitative analyses. 
These correlation coefficients are listed in Table \ref{tab:ddr_sa_lhc_corr}. 
To decide threshold values for correlation coefficient descriptors, we here compromise between threshold values suggested by other authors (listed in table  \ref{tab:sa_corrcoeff_thresh}) whilst taking into account that we are only regarding parameter values within `plausible' ranges. 
With this as a guide, we here decide to refer to the linear input-output relationship as being `negligible' for $\gamma$, where the obtained correlation coefficients are 0.05 and 0.12 for $X^1$ and $X^2$ respectively. 
We further say that for $\Pi_{D-S}$ the linear input-output relationship is positively/negatively weak for $X^1$ and $X^2$ respectively. 
For $EC_{50}$, however, the linear input-output relationship is moderately negative for $X^1$ and strongly positive for $X^2$. Clearly, we must be careful when choosing our $EC_{50}$ value in the model, as this highly influences the model output!

\begin{table}[H]
\begin{center}
\label{tab:ddr_sa_lhc_corr}
\begin{tabular}{|x{2.5cm}||x{1.2cm}|x{1.2cm}|x{1.2cm}|}
\hline
       \diaghead(-3,2){\hskip \hsize}{output}{input}& $\Pi_{D-S}$ & $EC_{50}$ & $\gamma$ \\ \hline
       $X^1$ &  0.19  & -0.59 & 0.05  \\
       $X^2$ & -0.24  & 0.84 & 0.12 \\
       \hline
\end{tabular}
\end{center}
\caption{The Pearson Product Moment Correlation Coefficients, computed in the Latin Hypercube Analysis, quantitatively describe input-output relationships.}
\end{table}


\newpage 

\section{Conclusion}	
This review is intended as a gentle, introductory review to three uncertainty and sensitivity analyses methods, namely, Consistency Analysis, Robustness Analysis and Latin Hypercube Analysis. Information on how to implement these methods in \textsf{MATLAB} are available in the Appendix. Alternatively, all methods discussed in this review can be implemented using the R-based software package Spartan, developed by Alden {\em et al.} \cite{Spartan2013}. In fact, many of the proceedings and conventions used in this review follow those suggested by Alden {\em et al.} in order to allow the reader to, as easily as possible, use Spartan if desired. 
Scrutinising mathematical models using uncertainty and sensitivity analyses methods is an important part in model development. In many applications, knowledge about a model's robustness is crucial \cite{Visser2014}. 
In the context of quantitative pharmacology, for example, a mathematical model may be used to guide preclinical or, ultimately, clinical proceedings. In such cases, understanding how confident we can be with model results, and how sensitive a model is to parameter perturbations, is of the utmost importance. 

\section*{Acknowledgements}
SH was supported by the Medical Research Council [grant code MR/R017506/1] and Swansea University PhD Research Studentship. SS was supported by an STFC studentship under the DTP grant ST/N504464/1.

\section*{Appendix -- \textsf{MATLAB} code snippets}
\subsection*{Computing measure of stochastic superiority}
We here list two different \textsf{MATLAB} functions that can be used in order to compute the point estimate of the A-measure of stochastic superiority in the original form, $\hat{A} \in [0,1]$, and in the scaled form,  $\hat{\underline{A}} \in[0.5,1]$. 
The function \textsf{getA\_measure\_naive}, listed below, uses direct implementations of Equations \ref{eq:pointestimateA_sum} and \ref{eq:scaled_or_0p5to1} to compute and return values for $\hat{A}_{x_0,x_1}$ and $\underline{\hat{A}}_{x_0,x_1}$, given two input vectors $x_0$ and $x_1$. 
The function \textsf{getA\_measure} uses the built-in \textsf{MATLAB} function \textsf{ranksum} to do the same. 

\begin{lstlisting}[style=Matlab-editor,basicstyle=\mlttfamily]
function [A_measure, scaled_A_measure] = getA_measure(x0, x1)
    [p,h,stats] = ranksum(x0,x1);
    % Compute the A measure
    A_measure=(stats.ranksum/length(x0) - (length(x0)+1)/2)/length(x1);
    % Compute the scaled A measure
    scaled_A_measure=0.5+abs(0.5 -A_measure);
end
\end{lstlisting}

\begin{lstlisting}[style=Matlab-editor,basicstyle=\mlttfamily]
function [A_measure, scaled_A_measure] = getA_measure_naive(x0, x1)
    % Compute the A measure
    A_measure = 0;
    for i = 1:length(x0)
        for j = 1:length(x1)
            if(x0(i)>x1(j))
                A_measure = A_measure + 1; 
            elseif(x0(i)==x1(j))
                A_measure = A_measure + 0.5; 
            elseif(x0(i)<x1(j))
                A_measure = A_measure + 0; 
            end
        end
    end 
    A_measure = A_measure/(length(x0)*length(x1));
    % Compute the scaled A measure
    if(A_measure>=0.5)
        scaled_A_measure = A_measure;
    else
        scaled_A_measure = 1-A_measure;
    end
end
\end{lstlisting}

\subsection*{Creating boxplots}
The \textsf{MATLAB} function \textsf{boxplot} can be used to create boxplots. The input data in one column is represented by one box in the boxplot. For details regarding labeling and style alternatives, please see the \textsf{MATLAB} documentation \cite{MATLAB:2019b}.

\begin{lstlisting}[style=Matlab-editor,basicstyle=\mlttfamily]
boxplot(M);
\end{lstlisting}

\subsection*{Choosing Latin Hypercube Sampling points}
A Latin Hypercube Sampling matrix can be created using the \textsf{MATLAB} function \textsf{lhsdesign}, which returns a matrix of size $n \times q$, where $n$ denotes the number of samples to be tested, and $q$ denotes the number of input parameters to investigate (and thus perturb).

\begin{lstlisting}[style=Matlab-editor,basicstyle=\mlttfamily]
LHC_Matrix=lhsdesign(n,q)
\end{lstlisting}

Each row $i$, in the created matrix (here denoted \textsf{LHC\_Matrix}), corresponds to the $i$th sampling point. Each element ($i,j$) corresponds to the parameter value of the $j$th input parameter in sampling point $i$, where each parameter ranges between 0 and 1. 
For different criteria on how to chose the specific parameter values within each sampled interval, please refer to the \textsf{MATLAB} documentation \cite{MATLAB:2019b}. Sampling points can, for example, be chosen in a way that maximises the distance between sampling points in the $q$-dimensional sampling space.

\subsection*{Qualitative and Quantitative Latin Hypercube Sampling Analysis}
In order to qualitatively asses the correlation between an input parameter $p$, and an output response $X$, one can use the \textsf{MATLAB} function \textsf{scatter}. In the below listings, \textsf{p} and \textsf{X} are two data vectors.  

\begin{lstlisting}[style=Matlab-editor,basicstyle=\mlttfamily]
scatter(p,X)
\end{lstlisting}

Further, to quantify the linear correlation between $p$ and $X$, the \textsf{MATLAB} function \textsf{corrcoef} can be used to compute correlation coefficients.  

\begin{lstlisting}[style=Matlab-editor,basicstyle=\mlttfamily]
R=corrcoef(p, X);
\end{lstlisting}

\bibliography{manuscript}

\end{document}